\documentclass[10pt, conference, letterpaper, nofonttune]{IEEEtran}

\usepackage{url}
\usepackage[utf8]{inputenc}
\usepackage{xcolor}
\usepackage{amsmath}
\usepackage{xspace}
\usepackage{epsfig}
\usepackage{balance}
\usepackage{booktabs}
\usepackage{tabularx}
\usepackage[font=footnotesize]{subcaption}
\usepackage[font=footnotesize]{caption}
\usepackage{mathtools}
\usepackage[numbers,sort&compress]{natbib}
\usepackage{soul}
\usepackage{lipsum}
\usepackage{bm}

\usepackage[acronyms,nonumberlist,nopostdot,nomain,nogroupskip,acronymlists={hidden}]{glossaries}
\newglossary[algh]{hidden}{acrh}{acnh}{Hidden Acronyms}
\glsdisablehyper

\usepackage{tikz}
\usepackage{pgfplots}
\pgfplotsset{compat=newest}
\pgfplotsset{plot coordinates/math parser=false}
\newlength\fheight
\newlength\fwidth
\usetikzlibrary{plotmarks,patterns,decorations.pathreplacing,backgrounds,calc,arrows,arrows.meta,spy,matrix,scopes}
\usepgfplotslibrary{patchplots,groupplots}
\usepackage{tikzscale}
\usepackage[draft]{hyperref}

\newacronym{3gpp}{3GPP}{3rd Generation Partnership Project}
\newacronym{4g}{4G}{4th generation}
\newacronym{5g}{5G}{5th generation}
\newacronym{5gc}{5GC}{5G Core}
\newacronym{adc}{ADC}{Analog to Digital Converter}
\newacronym{aerpaw}{AERPAW}{Aerial Experimentation and Research Platform for Advanced Wireless}
\newacronym{ai}{AI}{Artificial Intelligence}
\newacronym{aimd}{AIMD}{Additive Increase Multiplicative Decrease}
\newacronym{am}{AM}{Acknowledged Mode}
\newacronym{amc}{AMC}{Adaptive Modulation and Coding}
\newacronym{amf}{AMF}{Access and Mobility Management Function}
\newacronym{aops}{AOPS}{Adaptive Order Prediction Scheduling}
\newacronym{api}{API}{Application Programming Interface}
\newacronym{apn}{APN}{Access Point Name}
\newacronym{aqm}{AQM}{Active Queue Management}
\newacronym{ausf}{AUSF}{Authentication Server Function}
\newacronym{avc}{AVC}{Advanced Video Coding}
\newacronym{awgn}{AGWN}{Additive White Gaussian Noise}
\newacronym{balia}{BALIA}{Balanced Link Adaptation Algorithm}
\newacronym{bbu}{BBU}{Base Band Unit}
\newacronym{bdp}{BDP}{Bandwidth-Delay Product}
\newacronym{ber}{BER}{Bit Error Rate}
\newacronym{bf}{BF}{Beamforming}
\newacronym{bler}{BLER}{Block Error Rate}
\newacronym{brr}{BRR}{Bayesian Ridge Regressor}
\newacronym{bsr}{BSR}{Buffer Status Report}
\newacronym{bs}{BS}{Base Station}
\newacronym{bss}{BSS}{Business Support System}
\newacronym{ca}{CA}{Carrier Aggregation}
\newacronym{caas}{CaaS}{Connectivity-as-a-Service}
\newacronym{cb}{CB}{Code Block}
\newacronym{cc}{CC}{Congestion Control}
\newacronym{ccid}{CCID}{Congestion Control ID}
\newacronym{cco}{CC}{Carrier Component}
\newacronym{cdd}{CDD}{Cyclic Delay Diversity}
\newacronym{cdf}{CDF}{Cumulative Distribution Function}
\newacronym{cdn}{CDN}{Content Distribution Network}
\newacronym{cir}{CIR}{Channel Impulse Response}
\newacronym{cn}{CN}{Core Network}
\newacronym{codel}{CoDel}{Controlled Delay Management}
\newacronym{comac}{COMAC}{Converged Multi-Access and Core}
\newacronym{cord}{CORD}{Central Office Re-architected as a Datacenter}
\newacronym{cornet}{CORNET}{COgnitive Radio NETwork}
\newacronym{cosmos}{COSMOS}{Cloud Enhanced Open Software Defined Mobile Wireless Testbed for City-Scale Deployment}
\newacronym{cots}{COTS}{Commercial Off-the-Shelf}
\newacronym{cp}{CP}{Control Plane}
\newacronym{cpu}{CPU}{Central Processing Unit}
\newacronym{cqi}{CQI}{Channel Quality Information}
\newacronym{cr}{CR}{Cognitive Radio}
\newacronym{cran}{CRAN}{Cloud \gls{ran}}
\newacronym{crs}{CRS}{Cell Reference Signal}
\newacronym{csi}{CSI}{Channel State Information}
\newacronym{csirs}{CSI-RS}{Channel State Information - Reference Signal}
\newacronym{cu}{CU}{Central Unit}
\newacronym{d2tcp}{D$^2$TCP}{Deadline-aware Data center TCP}
\newacronym{d3}{D$^3$}{Deadline-Driven Delivery}
\newacronym{dac}{DAC}{Digital to Analog Converter}
\newacronym{dag}{DAG}{Directed Acyclic Graph}
\newacronym{darpa}{DARPA}{Defense Advanced Research Projects Agency}
\newacronym{das}{DAS}{Distributed Antenna System}
\newacronym{dash}{DASH}{Dynamic Adaptive Streaming over HTTP}
\newacronym{dc}{DC}{Dual Connectivity}
\newacronym{dccp}{DCCP}{Datagram Congestion Control Protocol}
\newacronym{dce}{DCE}{Direct Code Execution}
\newacronym{dci}{DCI}{Downlink Control Information}
\newacronym{dcl}{DCL}{Dear Colleague Letter}
\newacronym{dctcp}{DCTCP}{Data Center TCP}
\newacronym{dl}{DL}{Downlink}
\newacronym{dmr}{DMR}{Deadline Miss Ratio}
\newacronym{dmrs}{DMRS}{DeModulation Reference Signal}
\newacronym{drlcc}{DRL-CC}{Deep Reinforcement Learning Congestion Control}
\newacronym{drs}{DRS}{Discovery Reference Signal}
\newacronym{du}{DU}{Distributed Unit}
\newacronym{e2e}{E2E}{end-to-end}
\newacronym{ecaas}{ECaaS}{Edge-Cloud-as-a-Service}
\newacronym{ecn}{ECN}{Explicit Congestion Notification}
\newacronym{edf}{EDF}{Earliest Deadline First}
\newacronym{embb}{eMBB}{Enhanced Mobile Broadband}
\newacronym{empower}{EMPOWER}{EMpowering transatlantic PlatfOrms for advanced WirEless Research}
\newacronym{enb}{eNB}{evolved Node Base}
\newacronym{endc}{EN-DC}{E-UTRAN-\gls{nr} \gls{dc}}
\newacronym{epc}{EPC}{Evolved Packet Core}
\newacronym{eps}{EPS}{Evolved Packet System}
\newacronym{es}{ES}{Edge Server}
\newacronym{etsi}{ETSI}{European Telecommunications Standards Institute}
\newacronym[firstplural=Estimated Times of Arrival (ETAs)]{eta}{ETA}{Estimated Time of Arrival}
\newacronym{eutran}{E-UTRAN}{Evolved Universal Terrestrial Access Network}
\newacronym{faas}{FaaS}{Function-as-a-Service}
\newacronym{fapi}{FAPI}{Functional Application Platform Interface}
\newacronym{fcc}{FCC}{Federal Communications Commission}
\newacronym{fdd}{FDD}{Frequency Division Duplexing}
\newacronym{fdm}{FDM}{Frequency Division Multiplexing}
\newacronym{fdma}{FDMA}{Frequency Division Multiple Access}
\newacronym{fed4fire}{FED4FIRE+}{Federation 4 Future Internet Research and Experimentation Plus}
\newacronym{fir}{FIR}{Finite Impulse Response}
\newacronym{fit}{FIT}{Future \acrlong{iot}}
\newacronym{fpga}{FPGA}{Field Programmable Gate Array}
\newacronym{fr2}{FR2}{Frequency Range 2}
\newacronym{fs}{FS}{Fast Switching}
\newacronym{fscc}{FSCC}{Flow Sharing Congestion Control}
\newacronym{ftp}{FTP}{File Transfer Protocol}
\newacronym{fw}{FW}{Flow Window}
\newacronym{ge}{GE}{Gaussian Elimination}
\newacronym{gnb}{gNB}{Next Generation Node Base}
\newacronym{gop}{GOP}{Group of Pictures}
\newacronym{gpr}{GPR}{Gaussian Process Regressor}
\newacronym{gpu}{GPU}{Graphics Processing Unit}
\newacronym{gtp}{GTP}{GPRS Tunneling Protocol}
\newacronym{gtpc}{GTP-C}{GPRS Tunnelling Protocol Control Plane}
\newacronym{gtpu}{GTP-U}{GPRS Tunnelling Protocol User Plane}
\newacronym{gtpv2c}{GTPv2-C}{\gls{gtp} v2 - Control}
\newacronym{gw}{GW}{Gateway}
\newacronym{harq}{HARQ}{Hybrid Automatic Repeat reQuest}
\newacronym{hetnet}{HetNet}{Heterogeneous Network}
\newacronym{hh}{HH}{Hard Handover}
\newacronym{hol}{HOL}{Head-of-Line}
\newacronym{hqf}{HQF}{Highest-quality-first}
\newacronym{hss}{HSS}{Home Subscription Server}
\newacronym{http}{HTTP}{HyperText Transfer Protocol}
\newacronym{ia}{IA}{Initial Access}
\newacronym{iab}{IAB}{Integrated Access and Backhaul}
\newacronym{ic}{IC}{Incident Command}
\newacronym{ietf}{IETF}{Internet Engineering Task Force}
\newacronym{imsi}{IMSI}{International Mobile Subscriber Identity}
\newacronym{imt}{IMT}{International Mobile Telecommunication}
\newacronym{iot}{IoT}{Internet of Things}
\newacronym{ip}{IP}{Internet Protocol}
\newacronym{itu}{ITU}{International Telecommunication Union}
\newacronym{kpi}{KPI}{Key Performance Indicator}
\newacronym{kvm}{KVM}{Kernel-based Virtual Machine}
\newacronym{los}{LOS}{Line-of-Sight}
\newacronym{lsm}{LSM}{Link-to-System Mapping}
\newacronym{lstm}{LSTM}{Long Short Term Memory}
\newacronym{lte}{LTE}{Long Term Evolution}
\newacronym{lxc}{LXC}{Linux Container}
\newacronym{m2m}{M2M}{Machine to Machine}
\newacronym{mac}{MAC}{Medium Access Control}
\newacronym{manet}{MANET}{Mobile Ad Hoc Network}
\newacronym{mano}{MANO}{Management and Orchestration}
\newacronym{mc}{MC}{Multi-Connectivity}
\newacronym{mcc}{MCC}{Mobile Cloud Computing}
\newacronym{mchem}{MCHEM}{Massive Channel Emulator}
\newacronym{mcs}{MCS}{Modulation and Coding Scheme}
\newacronym{mec}{MEC}{Multi-access Edge Computing}
\newacronym{mec2}{MEC}{Mobile Edge Cloud}
\newacronym{mfc}{MFC}{Mobile Fog Computing}
\newacronym{mi}{MI}{Mutual Information}
\newacronym{mib}{MIB}{Master Information Block}
\newacronym{miesm}{MIESM}{Mutual Information Based Effective SINR}
\newacronym{mimo}{MIMO}{Multiple Input, Multiple Output}
\newacronym{mgen}{MGEN}{Multi-Generator}
\newacronym{ml}{ML}{Machine Learning}
\newacronym{mlr}{MLR}{Maximum-local-rate}
\newacronym[plural=\gls{mme}s,firstplural=Mobility Management Entities (MMEs)]{mme}{MME}{Mobility Management Entity}
\newacronym{mmtc}{mMTC}{Massive Machine-Type Communications}
\newacronym{mmwave}{mmWave}{millimeter wave}
\newacronym{mpdccp}{MP-DCCP}{Multipath Datagram Congestion Control Protocol}
\newacronym{mptcp}{MPTCP}{Multipath TCP}
\newacronym{mr}{MR}{Maximum Rate}
\newacronym{mrdc}{MR-DC}{Multi \gls{rat} \gls{dc}}
\newacronym{mse}{MSE}{Mean Square Error}
\newacronym{mss}{MSS}{Maximum Segment Size}
\newacronym{mt}{MT}{Mobile Termination}
\newacronym{mtd}{MTD}{Machine-Type Device}
\newacronym{mtu}{MTU}{Maximum Transmission Unit}
\newacronym{mumimo}{MU-MIMO}{Multi-user \gls{mimo}}
\newacronym{mvno}{MVNO}{Mobile Virtual Network Operator}
\newacronym{nalu}{NALU}{Network Abstraction Layer Unit}
\newacronym{nas}{NAS}{Network Attached Storage}
\newacronym{nbiot}{NB-IoT}{Narrow Band IoT}
\newacronym{nfv}{NFV}{Network Function Virtualization}
\newacronym{nfvi}{NFVI}{Network Function Virtualization Infrastructure}
\newacronym{nic}{NIC}{Network Interface Card}
\newacronym{nlos}{NLOS}{Non-Line-of-Sight}
\newacronym{now}{NOW}{Non Overlapping Window}
\newacronym{nrdz}{NRDZ}{National Radio Dynamic Zone}
\newacronym{nsf}{NSF}{National Science Foundation}
\newacronym{nsm}{NSM}{Network Service Mesh}
\newacronym[type=hidden]{nr}{NR}{New Radio}
\newacronym{nrf}{NRF}{Network Repository Function}
\newacronym{nsa}{NSA}{Non Stand Alone}
\newacronym{nse}{NSE}{Network Slicing Engine}
\newacronym{nssf}{NSSF}{Network Slice Selection Function}
\newacronym{ntp}{NTP}{Network Time Protocol}
\newacronym{o2i}{O2I}{Outdoor to Indoor}
\newacronym{oai}{OAI}{OpenAirInterface}
\newacronym{oaicn}{OAI-CN}{\gls{oai} \acrlong{cn}}
\newacronym{oairan}{OAI-RAN}{\acrlong{oai} \acrlong{ran}}
\newacronym{oam}{OAM}{Operations, Administration and Maintenance}
\newacronym{ofdm}{OFDM}{Orthogonal Frequency Division Multiplexing}
\newacronym{olia}{OLIA}{Opportunistic Linked Increase Algorithm}
\newacronym{omec}{OMEC}{Open Mobile Evolved Core}
\newacronym{onap}{ONAP}{Open Network Automation Platform}
\newacronym{onf}{ONF}{Open Networking Foundation}
\newacronym{onos}{ONOS}{Open Networking Operating System}
\newacronym{oom}{OOM}{\gls{onap} Operations Manager}
\newacronym{opnfv}{OPNFV}{Open Platform for \gls{nfv}}
\newacronym[type=hidden]{oran}{O-RAN}{Open \gls{ran}}
\newacronym{orbit}{ORBIT}{Open-Access Research Testbed for Next-Generation Wireless Networks}
\newacronym{os}{OS}{Operating System}
\newacronym{oss}{OSS}{Operations Support System}
\newacronym{pa}{PA}{Position-aware}
\newacronym{pase}{PASE}{Prioritization, Arbitration, and Self-adjusting Endpoints}
\newacronym{pawr}{PAWR}{Platforms for Advanced Wireless Research}
\newacronym{pbch}{PBCH}{Physical Broadcast Channel}
\newacronym{pcef}{PCEF}{Policy and Charging Enforcement Function}
\newacronym{pcfich}{PCFICH}{Physical Control Format Indicator Channel}
\newacronym{pcrf}{PCRF}{Policy and Charging Rules Function}
\newacronym{pdcch}{PDCCH}{Physical Downlink Control Channel}
\newacronym{pdcp}{PDCP}{Packet Data Convergence Protocol}
\newacronym{pdsch}{PDSCH}{Physical Downlink Shared Channel}
\newacronym{pdu}{PDU}{Packet Data Unit}
\newacronym{pf}{PF}{Proportional Fair}
\newacronym{pgw}{PGW}{Packet Gateway}
\newacronym{phich}{PHICH}{Physical Hybrid ARQ Indicator Channel}
\newacronym{phy}{PHY}{Physical}
\newacronym{pmch}{PMCH}{Physical Multicast Channel}
\newacronym{pmi}{PMI}{Precoding Matrix Indicators}
\newacronym{powder}{POWDER}{Platform for Open Wireless Data-driven Experimental Research}
\newacronym{ppo}{PPO}{Proximal Policy Optimization}
\newacronym{ppp}{PPP}{Poisson Point Process}
\newacronym{prach}{PRACH}{Physical Random Access Channel}
\newacronym{prb}{PRB}{Physical Resource Block}
\newacronym{psnr}{PSNR}{Peak Signal to Noise Ratio}
\newacronym{pss}{PSS}{Primary Synchronization Signal}
\newacronym{pucch}{PUCCH}{Physical Uplink Control Channel}
\newacronym{pusch}{PUSCH}{Physical Uplink Shared Channel}
\newacronym{qam}{QAM}{Quadrature Amplitude Modulation}
\newacronym{qci}{QCI}{\gls{qos} Class Identifier}
\newacronym{qoe}{QoE}{Quality of Experience}
\newacronym{qos}{QoS}{Quality of Service}
\newacronym{quic}{QUIC}{Quick UDP Internet Connections}
\newacronym{rach}{RACH}{Random Access Channel}
\newacronym{ran}{RAN}{Radio Access Network}
\newacronym[firstplural=Radio Access Technologies (RATs)]{rat}{RAT}{Radio Access Technology}
\newacronym{rcn}{RCN}{Research Coordination Network}
\newacronym{rec}{REC}{Radio Edge Cloud}
\newacronym{red}{RED}{Random Early Detection}
\newacronym{renew}{RENEW}{Reconfigurable Eco-system for Next-generation End-to-end Wireless}
\newacronym{rf}{RF}{Radio Frequency}
\newacronym{rfc}{RFC}{Request for Comments}
\newacronym{rfr}{RFR}{Random Forest Regressor}
\newacronym{ric}{RIC}{\gls{ran} Intelligent Controller}
\newacronym{rlc}{RLC}{Radio Link Control}
\newacronym{rlf}{RLF}{Radio Link Failure}
\newacronym{rlnc}{RLNC}{Random Linear Network Coding}
\newacronym{rmse}{RMSE}{Root Mean Squared Error}
\newacronym{rnis}{RNIS}{Radio Network Information Service}
\newacronym{rr}{RR}{Round Robin}
\newacronym{rrc}{RRC}{Radio Resource Control}
\newacronym{rrm}{RRM}{Radio Resource Management}
\newacronym{rru}{RRU}{Remote Radio Unit}
\newacronym{rs}{RS}{Remote Server}
\newacronym{rsrp}{RSRP}{Reference Signal Received Power}
\newacronym{rsrq}{RSRQ}{Reference Signal Received Quality}
\newacronym{rss}{RSS}{Received Signal Strength}
\newacronym{rssi}{RSSI}{Received Signal Strength Indicator}
\newacronym{rtt}{RTT}{Round Trip Time}
\newacronym{ru}{RU}{Radio Unit}
\newacronym{rw}{RW}{Receive Window}
\newacronym{rx}{RX}{Receiver}
\newacronym{s1ap}{S1AP}{S1 Application Protocol}
\newacronym{sa}{SA}{standalone}
\newacronym{sack}{SACK}{Selective Acknowledgment}
\newacronym{sap}{SAP}{Service Access Point}
\newacronym{sc2}{SC2}{Spectrum Collaboration Challenge}
\newacronym{scef}{SCEF}{Service Capability Exposure Function}
\newacronym{sch}{SCH}{Secondary Cell Handover}
\newacronym{scoot}{SCOOT}{Split Cycle Offset Optimization Technique}
\newacronym{sctp}{SCTP}{Stream Control Transmission Protocol}
\newacronym{sdap}{SDAP}{Service Data Adaptation Protocol}
\newacronym{sdk}{SDK}{Software Development Kit}
\newacronym{sdm}{SDM}{Space Division Multiplexing}
\newacronym{sdma}{SDMA}{Spatial Division Multiple Access}
\newacronym{sdn}{SDN}{Software-defined Networking}
\newacronym{sdr}{SDR}{Software-defined Radio}
\newacronym{seba}{SEBA}{SDN-Enabled Broadband Access}
\newacronym{sgsn}{SGSN}{Serving GPRS Support Node}
\newacronym{sgw}{SGW}{Service Gateway}
\newacronym{si}{SI}{Study Item}
\newacronym{sib}{SIB}{Secondary Information Block}
\newacronym{sinr}{SINR}{Signal to Interference plus Noise Ratio}
\newacronym{sip}{SIP}{Session Initiation Protocol}
\newacronym{siso}{SISO}{Single Input, Single Output}
\newacronym{sla}{SLA}{Service Level Agreement}
\newacronym{sm}{SM}{Saturation Mode}
\newacronym{smf}{SMF}{Session Management Function}
\newacronym{smo}{SMO}{Service Management and Orchestration}
\newacronym{sms}{SMS}{Short Message Service}
\newacronym{smsgmsc}{SMS-GMSC}{\gls{sms}-Gateway}
\newacronym{snr}{SNR}{Signal-to-Noise-Ratio}
\newacronym{son}{SON}{Self-Organizing Network}
\newacronym{sptcp}{SPTCP}{Single Path TCP}
\newacronym{srb}{SRB}{Service Radio Bearer}
\newacronym{srn}{SRN}{Standard Radio Node}
\newacronym{srs}{SRS}{Sounding Reference Signal}
\newacronym{ss}{SS}{Synchronization Signal}
\newacronym{sss}{SSS}{Secondary Synchronization Signal}
\newacronym{st}{ST}{Spanning Tree}
\newacronym{svc}{SVC}{Scalable Video Coding}
\newacronym{tb}{TB}{Transport Block}
\newacronym{tcp}{TCP}{Transmission Control Protocol}
\newacronym{tdd}{TDD}{Time Division Duplexing}
\newacronym{tdm}{TDM}{Time Division Multiplexing}
\newacronym{tdma}{TDMA}{Time Division Multiple Access}
\newacronym{tfl}{TfL}{Transport for London}
\newacronym{tfrc}{TFRC}{TCP-Friendly Rate Control}
\newacronym{tft}{TFT}{Traffic Flow Template}
\newacronym{tgen}{TGEN}{Traffic Generator}
\newacronym{tip}{TIP}{Telecom Infra Project}
\newacronym{tm}{TM}{Transparent Mode}
\newacronym{to}{TO}{Telco Operator}
\newacronym{tr}{TR}{Technical Report}
\newacronym{trp}{TRP}{Transmitter Receiver Pair}
\newacronym{ts}{TS}{Technical Specification}
\newacronym{tti}{TTI}{Transmission Time Interval}
\newacronym{ttt}{TTT}{Time-to-Trigger}
\newacronym{tx}{TX}{Transmitter}
\newacronym{uas}{UAS}{Unmanned Aerial System}
\newacronym{uav}{UAV}{Unmanned Aerial Vehicle}
\newacronym{udm}{UDM}{Unified Data Management}
\newacronym{udp}{UDP}{User Datagram Protocol}
\newacronym{udr}{UDR}{Unified Data Repository}
\newacronym{ue}{UE}{User Equipment}
\newacronym{uhd}{UHD}{\gls{usrp} Hardware Driver}
\newacronym{ul}{UL}{Uplink}
\newacronym{um}{UM}{Unacknowledged Mode}
\newacronym{uml}{UML}{Unified Modeling Language}
\newacronym{upa}{UPA}{Uniform Planar Array}
\newacronym{upf}{UPF}{User Plane Function}
\newacronym{urllc}{URLLC}{Ultra Reliable and Low Latency Communication}
\newacronym{usa}{U.S.}{United States}
\newacronym{usim}{USIM}{Universal Subscriber Identity Module}
\newacronym{usrp}{USRP}{Universal Software Radio Peripheral}
\newacronym{utc}{UTC}{Urban Traffic Control}
\newacronym{vim}{VIM}{Virtualization Infrastructure Manager}
\newacronym{vm}{VM}{Virtual Machine}
\newacronym{vnf}{VNF}{Virtual Network Function}
\newacronym{volte}{VoLTE}{Voice over \gls{lte}}
\newacronym{voltha}{VOLTHA}{Virtual OLT HArdware Abstraction}
\newacronym{vr}{VR}{Virtual Reality}
\newacronym{vran}{vRAN}{Virtualized \gls{ran}}
\newacronym{vss}{VSS}{Video Streaming Server}
\newacronym{wbf}{WBF}{Wired Bias Function}
\newacronym{wf}{WF}{Wired-first}
\newacronym{wlan}{WLAN}{Wireless Local Area Network}
\newacronym{osm}{OSM}{Open Source \gls{nfv} Management and Orchestration}
\newacronym{pnf}{PNF}{Physical Network Function}
\newacronym{drl}{DRL}{Deep Reinforcement Learning}
\newacronym{mtc}{MTC}{Machine-type Communications}
\tikzstyle{startstop} = [rectangle, rounded corners, minimum width=2cm, minimum height=0.5cm,text centered, draw=black]
\tikzstyle{io} = [trapezium, trapezium left angle=70, trapezium right angle=110, minimum width=3cm, minimum height=1cm, text centered, draw=black]
\tikzstyle{process} = [rectangle, minimum width=2cm, minimum height=0.5cm, text centered, draw=black, alignb=center]
\tikzstyle{decision} = [ellipse, minimum width=2cm, minimum height=1cm, text centered, draw=black]
\tikzstyle{arrow} = [thick,<->,>=stealth]
\tikzstyle{line} = [thick,>=stealth]
\tikzstyle{darrow} = [thick,<->,>=stealth,dashed]
\tikzstyle{sarrow} = [thick,->,>=stealth]
\tikzstyle{larrow} = [line width=0.1mm,dashdotted,->,>=stealth]
\tikzstyle{llarrow} = [line width=0.1mm,->,>=stealth]

\makeatletter
\def\grd@save@target#1{%
  \def\grd@target{#1}}
\def\grd@save@start#1{%
  \def\grd@start{#1}}
\tikzset{
  grid with coordinates/.style={
    to path={%
      \pgfextra{%
        \edef\grd@@target{(\tikztotarget)}%
        \tikz@scan@one@point\grd@save@target\grd@@target\relax
        \edef\grd@@start{(\tikztostart)}%
        \tikz@scan@one@point\grd@save@start\grd@@start\relax
        \draw[minor help lines] (\tikztostart) grid (\tikztotarget);
        \draw[major help lines] (\tikztostart) grid (\tikztotarget);
        \grd@start
        \pgfmathsetmacro{\grd@xa}{\the\pgf@x/1cm}
        \pgfmathsetmacro{\grd@ya}{\the\pgf@y/1cm}
        \grd@target
        \pgfmathsetmacro{\grd@xb}{\the\pgf@x/1cm}
        \pgfmathsetmacro{\grd@yb}{\the\pgf@y/1cm}
        \pgfmathsetmacro{\grd@xc}{\grd@xa + \pgfkeysvalueof{/tikz/grid with coordinates/major step x}}
        \pgfmathsetmacro{\grd@yc}{\grd@ya + \pgfkeysvalueof{/tikz/grid with coordinates/major step y}}
        \foreach \x in {\grd@xa,\grd@xc,...,\grd@xb}
        \node[anchor=north] at (\x,\grd@ya) {\pgfmathprintnumber{\x}};
        \foreach \y in {\grd@ya,\grd@yc,...,\grd@yb}
        \node[anchor=east] at (\grd@xa,\y) {\pgfmathprintnumber{\y}};
      }
    }
  },
  minor help lines/.style={
    help lines,
    gray,
    line cap =round,
    xstep=\pgfkeysvalueof{/tikz/grid with coordinates/minor step x},
    ystep=\pgfkeysvalueof{/tikz/grid with coordinates/minor step y}
  },
  major help lines/.style={
    help lines,
    line cap =round,
    line width=\pgfkeysvalueof{/tikz/grid with coordinates/major line width},
    xstep=\pgfkeysvalueof{/tikz/grid with coordinates/major step x},
    ystep=\pgfkeysvalueof{/tikz/grid with coordinates/major step y}
  },
  grid with coordinates/.cd,
  minor step x/.initial=.5,
  minor step y/.initial=.2,
  major step x/.initial=1,
  major step y/.initial=1,
  major line width/.initial=1pt,
}
\makeatother

\newif\ifexttikz
\exttikztrue

\ifexttikz
  \usetikzlibrary{external}
\fi

\newif\ifoverleaf
\overleaftrue

\begin{document}
\bstctlcite{BSTcontrol}

\title{Colosseum: Large-Scale Wireless Experimentation Through Hardware-in-the-Loop Network Emulation\vspace{-0.4cm}}

\author{\IEEEauthorblockN{Leonardo Bonati,\IEEEauthorrefmark{1} 
Pedram Johari,\IEEEauthorrefmark{1}
Michele Polese,\IEEEauthorrefmark{1}
Salvatore D'Oro,\IEEEauthorrefmark{1}
Subhramoy Mohanti,\IEEEauthorrefmark{1}\\
Miead Tehrani-Moayyed,\IEEEauthorrefmark{1}
Davide Villa,\IEEEauthorrefmark{1}
Shweta Shrivastava,\IEEEauthorrefmark{1}
Chinenye Tassie,\IEEEauthorrefmark{1}
Kurt Yoder,\IEEEauthorrefmark{2}\\
Ajeet Bagga,\IEEEauthorrefmark{3}
Paresh Patel,\IEEEauthorrefmark{3}
Ventz Petkov,\IEEEauthorrefmark{3}
Michael Seltser,\IEEEauthorrefmark{3}
Francesco Restuccia,\IEEEauthorrefmark{1}\IEEEauthorrefmark{4}\\
Abhimanyu Gosain,\IEEEauthorrefmark{1}
Kaushik R.\ Chowdhury,\IEEEauthorrefmark{1}
Stefano Basagni,\IEEEauthorrefmark{1}
Tommaso Melodia\IEEEauthorrefmark{1}}
\IEEEauthorblockA{
\IEEEauthorrefmark{1}Institute for the Wireless Internet of Things, Northeastern University, Boston, MA, U.S.A.\\
\IEEEauthorrefmark{4}Roux Institute, Northeastern University, Portland, ME, U.S.A.\\
\IEEEauthorrefmark{2}Greenfly SAU LLC, Dunn Loring, VA, U.S.A.\hspace{0.25cm}
\IEEEauthorrefmark{3}Cerbo IO LLC, Burlington, MA, U.S.A.
\vspace{-1.5cm}}\\
\thanks{Contact: melodia@northeastern.edu. This work was partially supported by the U.S.\ National Science Foundation under grant CNS-1925601.}
}

\IEEEoverridecommandlockouts

\maketitle

\glsunset{fpga}
\glsunset{nr}
\glsunset{usrp}

\begin{abstract}
Colosseum is an open-access and publicly-available large-scale wireless testbed for experimental research via virtualized and softwarized waveforms and protocol stacks on a fully programmable, ``white-box'' platform.
Through 256~state-of-the-art software-defined radios and a massive channel emulator core, Colosseum can model virtually any scenario, enabling the design, development and testing of solutions at scale in a variety of deployments and channel conditions.
These Colosseum \emph{radio-frequency scenarios} are reproduced through high-fidelity FPGA-based emulation with finite-impulse response filters.
Filters model the taps of desired wireless channels and apply them to the signals generated by the radio nodes, faithfully mimicking the conditions of real-world wireless environments.
In this paper, we introduce Colosseum as a testbed that is \textit{for the first time} open to the research community.
We describe the architecture of Colosseum and its experimentation and emulation capabilities.
We then demonstrate the effectiveness of Colosseum for experimental research at scale through exemplary use cases including prevailing wireless technologies (e.g., cellular and Wi-Fi) in spectrum sharing and unmanned aerial vehicle scenarios.
A roadmap for Colosseum future updates concludes the paper.
\end{abstract}


\begin{picture}(0,0)(10,-430)
\put(0,0){
\put(0,10){\footnotesize This paper has been accepted for publication on IEEE International Symposium on Dynamic Spectrum Access Networks (DySPAN) 2021.}
\put(0,0){\tiny \copyright 2021 IEEE. Personal use of this material is permitted. Permission from IEEE must be obtained for all other uses, in any current or future media including reprinting/republishing}
\put(0,-5){\tiny this material for advertising or promotional purposes, creating new collective works, for resale or redistribution to servers or lists, or reuse of any copyrighted component of this work in other works.}
\put(0,-20){\scriptsize }}
\end{picture}

\glsresetall
\glsunset{fpga}
\glsunset{nr}
\glsunset{usrp}


\section{Introduction}

Wireless technologies are undergoing advances at a pace never seen before.
%
The \gls{5g} of cellular networks is evolving toward open and disaggregated deployments in a way never experienced by previous generations~\cite{bonati2020open}; 
Wi-Fi standards are multiplying and now occupy new portions of the electromagnetic spectrum~\cite{naik2021coexistance};
applications of wireless networking include new verticals, ranging from those involving \glspl{uav}~\cite{bertizzolo2020swarmcontrol,BuczekBBM21} to telemedicine~\cite{santagati2017softwaredefined}.
%

The wireless research community has increasingly recognized the need for programmable, large-scale experimental testbeds to validate new enabling technologies and application scenarios in realistic environments, at scale, and with a diversity of traffic, topology, and channel conditions. Notoriously, however, high-fidelity testbeds are not easily accessible, scale is often limited, and costs are high. As a result, innovative solutions in wireless are often  tested either through numerical network simulations (that often lack fidelity, especially as concerns modeling of hardware constraints and imperfections, or of the effects of interference and channel variations on applications), or in small laboratory setups (which are able to capture to some extent the characteristics of hardware devices of commercial deployments but not the scale of such deployments or real propagation environments and interference conditions). 

\begin{figure*}[t]
    \centering
    \setlength\belowcaptionskip{-10pt}
    \includegraphics[width=\textwidth]{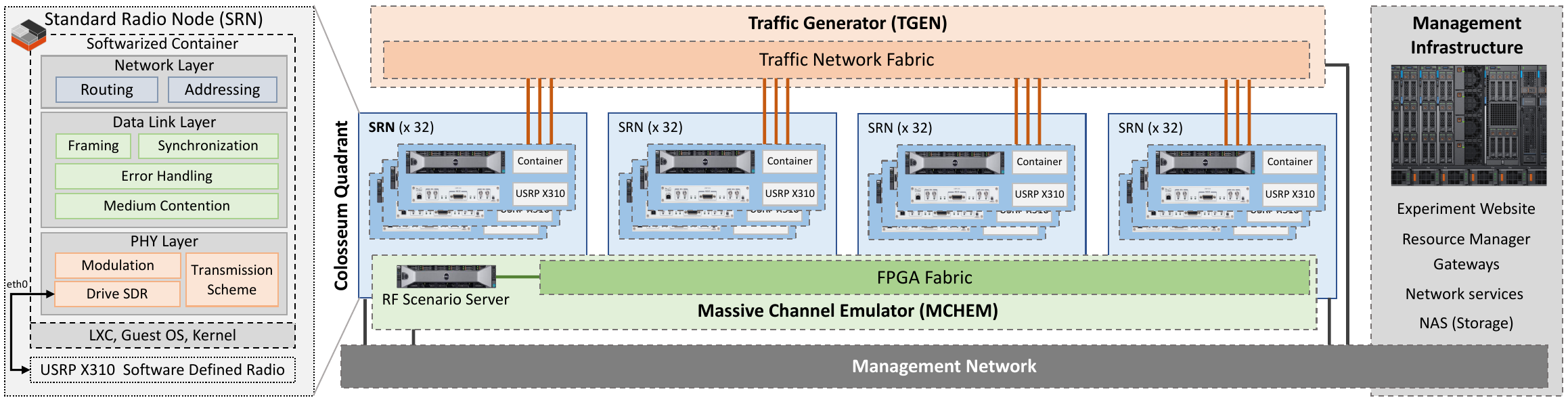}
    \caption{Colosseum architecture.}
    \label{fig:architecture}
\end{figure*}

To bridge this gap, a number of high-profile programs have attempted to develop shared, community-driven, and remotely-accessible experimental wireless platforms. A common goal of these initiatives is to create a widely-accepted and validated ecosystem for repeatable and reproducible experimentation in the wireless networking field. For example, in the United States, the \gls{nsf} has spearheaded the \gls{pawr} program. The goal of \gls{pawr} is to develop four city-scale programmable wireless platforms open to the research community at large. Among these, 
POWDER-RENEW focuses on developing a sub-$6\:\mathrm{GHz}$ programmable cellular architecture~\cite{breen2021powder} with massive MIMO capabilities~\cite{doostmohammady2021good}; COSMOS focuses on a combination of mmWave links~\cite{kohli2021openaccess} with low-latency programmable optical backhauls;  AERPAW focuses on the intersection between \gls{uav} verticals and cellular applications~\cite{panicker2021aerpaw}; while ARA will be focused on wireless rural scenarios~\cite{zhang2021ara}.

Similar efforts are being carried out in Europe by the 5GENESIS project~\cite{koumaras20185genesis}, which encompasses testbeds comprising an edge-computing \gls{nfv}-enabled radio infrastructure, orchestration and management frameworks, satellite communications,
and ultra-dense network deployments. Although the testbeds of this program complement each other in offering the tools to evaluate several aspects of the 5G ecosystem, they ignore non-cellular wireless access technologies. Smaller-scale indoor testbeds include the Drexel Grid~\cite{dandekar2019grid}, ORBIT~\cite{kohli2021openaccess} and Arena~\cite{bertizzolo2020arena}. The Drexel Grid couples physical \glspl{sdr} with virtual nodes capable of emulating custom channel conditions. ORBIT comprises a grid of \glspl{sdr} and compute nodes equipped with transceivers for heterogeneous wireless technologies (e.g., Bluetooth, ZigBee and LTE). Finally, Arena mounts a ceiling grid of 64 \gls{sdr}-driven antennas connected with edge compute and \gls{ai} capabilities. While these testbeds provide a good representation of indoor wireless propagation environments, their scale can hardly capture the dynamics of large real-world deployments.

In this paper, we present Colosseum---the world's largest wireless network emulator with hardware in-the-loop---as a platform that is \textit{for the first time} available to the research community.\footnote{\url{https://www.colosseum.net}}
Originally built by the \gls{darpa} and by the Johns Hopkins University Applied Physics Laboratory to support the \gls{sc2}~\cite{darpa_sc2, tilghman2019will, yuan2019defense, coleman2019overview, freeman2019software, plummer2019development, mok2019resource, white2019standard, yim2019incumbent, curtis2019traffic, barcklow2019radio}, Colosseum is being expanded and operated by the Institute for the Wireless Internet of Things\footnote{\url{https://www.northeastern.edu/wiot}} at Northeastern University through an \gls{nsf} grant, which also made it publicly available to the research community.
With its 256~\glspl{sdr} and 128~remotely-accessible compute nodes and GPUs, Colosseum provides the capabilities to test full-protocol stack solutions at scale with real hardware devices and in emulated---yet realistic---environments with complex channel interactions (e.g., path loss, fading, multipath).
%
Besides its experimentation capabilities, Colosseum can be used as an \gls{ai} playground and wireless data factory to create large-scale datasets and train/test solutions in a safe and controlled environment~\cite{bonati2021scope,bonati2021intelligence}.
%
We provide an overview of the architectural components and emulation capabilities of Colosseum in Sections~\ref{sec:architecture} and~\ref{sec:scenarios}. We show how to use Colosseum in Section~\ref{sec:how_to_use}, including practical use cases in Section~\ref{sec:usecases}. Finally, we describe the planned extensions to Colosseum in Section~\ref{sec:evolution} and we draw our conclusions in Section~\ref{sec:conclusions}.


\section{Colosseum Architecture}
\label{sec:architecture}

A high-level representation of the architecture of Colosseum is shown in Fig.~\ref{fig:architecture}. Colosseum comprises 128~\glspl{srn}, the \gls{mchem}, the \gls{rf} scenario server, the \gls{tgen}, and the management infrastructure. The \glspl{srn}, which can be controlled remotely to perform experiments, are divided in four quadrants and are synchronized in time and frequency through hierarchical OctoClock clock distributors.
Each \gls{srn} is a state-of-the-art server with 48-core~Intel Xeon E5-2650 CPUs and an NVIDIA Tesla K40m GPU, and drives an NI/Ettus \gls{usrp} X310.
Each X310 is equipped with two UBX-160 daughterboards that operate between $10\:\mathrm{MHz}$ and $6\:\mathrm{GHz}$.
%
Colosseum allows multiple concurrent users to automatically deploy softwarized containers---implemented via \glspl{lxc}---on the bare-metal \glspl{srn} (Fig.~\ref{fig:architecture}, left). Containers can be used to run a variety of protocol stacks and to control parameters and configurations at different layers of these stacks.

\begin{figure}[t]
    \centering
    \includegraphics[width=\columnwidth]{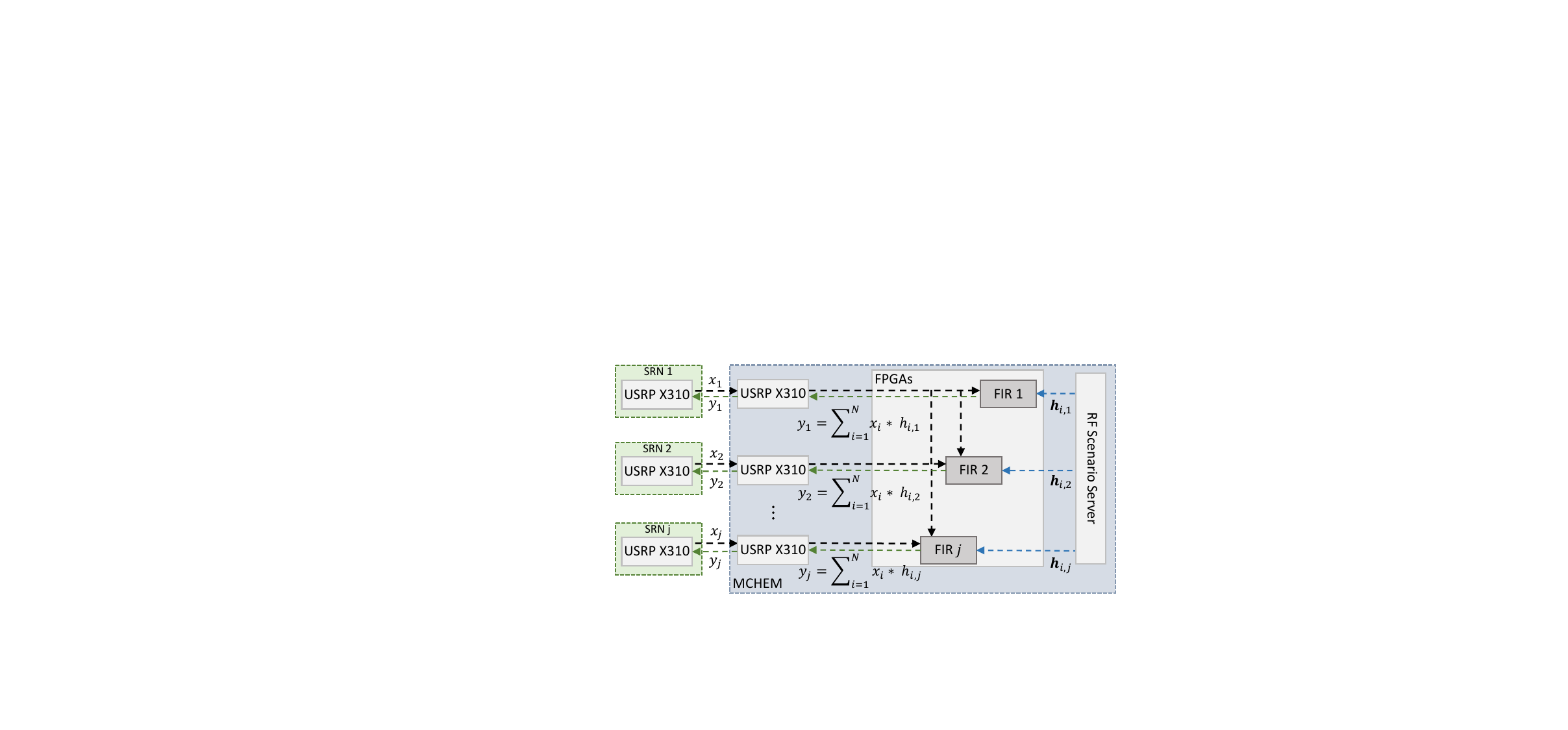}
    \caption{FPGA-based RF scenario emulation in Colosseum.}
    \label{fig:mchem_diagram}
    \vspace{-10pt}
\end{figure}

\gls{mchem} is tasked with the channel emulation in Colosseum.
It comprises four interconnected quadrants, each with 4~NI ATCA 3671~\gls{fpga} modules and 16~Virtex-7 690T \glspl{fpga}, and it drives an array of 128~\glspl{usrp} X310 interconnected in a one-to-one fashion to the \glspl{usrp} of the \glspl{srn} (see Fig.~\ref{fig:mchem_diagram}).
%
%

When an \gls{rf} transmission occurs in Colosseum, the signals generated by the \glspl{usrp} at the \gls{srn} side (e.g., signal $x_1$ in Fig.~\ref{fig:mchem_diagram}) get transmitted to the corresponding USRPs X310 of \gls{mchem}, which perform \gls{rf} to baseband and analog-to-digital conversions. The digital signals are then forwarded to the \glspl{fpga} of \gls{mchem} that process them through \gls{fir} filters.
Filters are composed of 512~pre-computed complex-valued taps that capture the characteristics of the channel, i.e., the \gls{cir}, between any pair of \glspl{srn}.\footnote{Due to the high computational complexity required to generate scenarios, and to the large space to store them, only 4~channel taps are non-zero-valued.}
As depicted in Fig.~\ref{fig:mchem_diagram}, the \gls{fir} filters load the vector $\bm{h}_{i,j}$ of the 512-tap \glspl{cir} among receive node $j$ and every transmit node $i$, with $i, j \in \{1,...,N\}$ set of \glspl{srn} of a specific experiment.
These channel taps are then applied to signal $x_i$ through a convolution operation. The resulting signal $y_{j} = \sum_{i=1}^{N} x_i \ast h_{i,j}$ (i.e., the transmitted signals $x_i$ processed with the \gls{cir} of the emulated channel between nodes $i$ and $j$) is then sent to \gls{srn} $j$. In this way, Colosseum aggregates all the received signals after passing them through the corresponding \gls{cir} and transmits the results $y_j$
to all the \glspl{srn} of the experiment (and not only to the intended receiver), thus capturing and emulating effects of real wireless channels, such as the interference among nodes and superimposition of signals from multiple nodes~\cite{ashish2018scalable}.

%

The \gls{rf} scenario server (see Fig.~\ref{fig:architecture} and~\ref{fig:mchem_diagram}) maintains a catalog of the Colosseum \gls{rf} scenarios and feeds their channel taps to the channel emulator at run time. Scenarios make it possible to emulate effects of the wireless channel, including path loss and fading in terrains up to $1\:\mathrm{km^2}$
and with up to $80\:\mathrm{MHz}$ bandwidth. The modular architecture of \gls{mchem} and the independence of its \glspl{usrp} allow Colosseum to concurrently emulate different scenarios on different experiments so that multiple users can operate on the system at the same time. As we will discuss in Section~\ref{sec:how_to_use}, the specific scenario to run can be selected by the user
through a specialized control interface.

Similarly to how \gls{mchem} emulates the \gls{rf} scenarios, \gls{tgen}---based on the U.S.\ Naval Research Laboratory's \gls{mgen}~\cite{mgen}---takes care of emulating IP traffic flows between the \glspl{srn}.
The set of traffic flows that can be generated by \gls{tgen}, together with their characteristics (e.g., packet rate, size and distribution), constitutes a \textit{traffic scenario}.
%
Once a traffic scenario starts, packets are delivered to the \glspl{srn}, which handle them through the user-defined protocol stack (e.g., by transmitting them through a cellular of Wi-Fi stack, see Section~\ref{sec:usecases}).
%
%
Finally, the Colosseum management infrastructure hosts a variety of auxiliary services (Fig.~\ref{fig:architecture}). These include: (i)~the website through which the users reserve resources and start experiments; (ii) the resource manager, which allocates resources to the users; (iii) gateways for user and management access to Colosseum; (iv) a $900\:\mathrm{TB}$ \gls{nas} to store \gls{lxc} images, experiment data and logs, and (v) various network services, including services to provide time synchronization across the whole testbed.

\section{Experimental Scenarios in Colosseum}
\label{sec:scenarios}

Experimental scenarios, i.e., \gls{rf} and traffic scenarios, are at the core of Colosseum. \gls{rf} scenarios enable the emulation of up to 65,535 wireless channels in diverse environments
with an emulation area up to $1\:\mathrm{km^2}$. They give users full control over the wireless environment, allowing them to capture and reproduce the desired channel effects while enabling reproducibility and repeatability of experiments at scale.
This emulation is carried out by \gls{mchem} through \gls{fpga}-based \gls{fir} filters, which capture and reproduce the \gls{cir} of the emulated wireless environment (see Section~\ref{sec:architecture}).
The \gls{cir} can be derived in many different ways: for instance, it can be computed through well-known channel modeling equations, performing field measurements~\cite{polese2020experimental}, or leveraging high-accuracy software tools, such as ray-tracers~\cite{tehranimoayyed2021creating}.
In this way, scenarios not only model the channel between transmitter and receiver pairs, but they also model effects typical of the wireless propagation environment, such as interference and superposition of the signals generated by multiple nodes. This ultimately guarantees a \textit{high-fidelity} channel emulation, ensuring that the emulated environments are as close as possible to real-world deployments.

\subsection{Examples of Available Experimental Scenarios}
\label{sec:scenario-samples}

We now provide a sample of large-scale scenarios that are available on Colosseum at the time of this writing: the \textit{Alleys of Austin} and the \textit{SC2 Championship Event (SCE) Qualification} scenarios---developed by \gls{darpa} for \gls{sc2}---and a set of \textit{cellular scenarios} developed as part of~\cite{bonati2021scope}.

\textbf{\textit{Alleys of Austin.}}
This scenario is set in the urban area of downtown Austin, TX, and involves 50~nodes divided in 5~groups, or \textit{squads}. Every group consists of 9~pedestrian users moving in a row formation, and a \gls{uav} circling on top of them.
The center frequency of this scenario is set to $1\:\mathrm{GHz}$. The duration of the scenario is $15$\:minutes, divided in three stages of $5$\:minutes each.
In the first stage, nodes exchange voice traffic.
In the second stage, they transmit images and videos in addition to the voice data of the previous phase.
Finally, in the third stage, the transmission rate of the nodes significantly increases.

\textbf{\textit{SCE Qualification.}}
This scenario concerns 10~nodes and has center frequency set to $1\:\mathrm{GHz}$. The duration of the scenario is $10$\:minutes.
The \gls{snr} among nodes is initially set to $20\:\mathrm{dB}$ and decreases by $5\:\mathrm{dB}$ every $2$\:minutes of the scenario.
Finally, in the last $2$\:minutes, the scenario center frequency shifts to $1.1\:\mathrm{GHz}$, and the \gls{snr} is reset to $20\:\mathrm{dB}$.
As for the traffic, nodes exchange UDP packets at constant bitrate.

\textbf{\textit{Cellular Scenarios.}}
Colosseum includes a number of scenarios designed for cellular networking experimentation developed as part of~\cite{bonati2021scope}. They involve 8~to 10~base stations serving four mobile users each. The locations of the base stations---derived from the OpenCelliD database~\cite{opencellid}---match those of real-world cellular deployments in Rome, Italy, in Boston, MA, and in the POWDER testbed~\cite{breen2021powder} of Salt Lake City, UT. The central frequency of these scenarios is set to $1\:\mathrm{GHz}$ and they have a duration of $10$\:minutes each. Users are deployed at different distances from the base stations, i.e., closer or farther from them, and they either move with pedestrian speed or remain still for the whole scenario. Finally, the uplink and downlink traffic profiles among users and base stations entail video packets for video streaming applications.

\subsection{Colosseum as a Wireless Data Factory}

Colosseum scale and emulation capabilities allow it to create large-scale datasets, namely acting as a wireless data factory. Its unique emulation capabilities---both in terms of \gls{rf} and traffic scenarios---allow users to configure an experiment once and then seamlessly run it in different environment, channel and traffic conditions.
Moreover, once set up, experiments can be run automatically in \textit{batches},
allowing users to collect data in bulk into wireless datasets (see Section~\ref{sec:how_to_use}).
%
These capabilities are fundamental when designing \gls{ml} algorithms, which need to be trained on large amounts of data and on different conditions to be able to generalize and adapt to unforeseen situations.

\section{Operational Modes of Colosseum}
\label{sec:how_to_use}

Colosseum allows users to operate in either (i) \textit{interactive mode}, in which users operate the \glspl{srn} manually, or (ii) \textit{batch mode}, in which experiments are carried out automatically.

\begin{figure}[ht]
\setlength\abovecaptionskip{0pt}
    \centering
    \includegraphics[width=0.9\columnwidth]{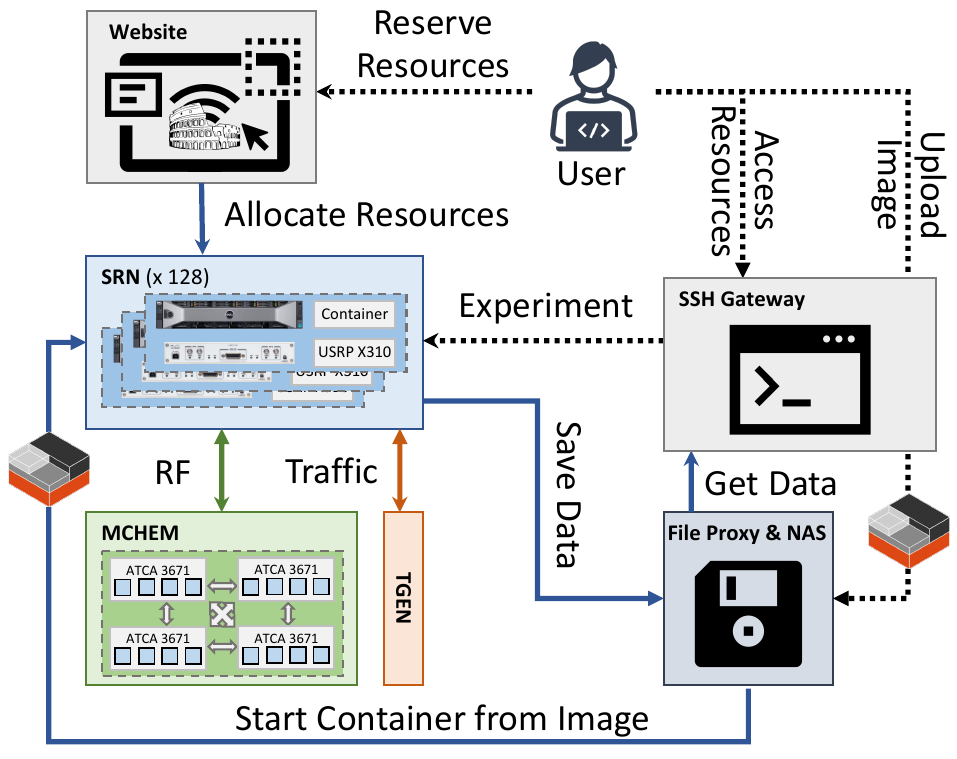}
    \caption{Workflow of a Colosseum interactive experiment.}
    \label{fig:access_diagram}
\end{figure}

\textbf{\textit{Interactive Mode.}}
The workflow of an interactive experiment in Colosseum is shown in Fig.~\ref{fig:access_diagram}.
As a first step, the users reserve a number of \glspl{srn} through Colosseum website. In this step, they can specify which \gls{lxc} image to load on each \gls{srn} (i.e., either one of Colosseum-provided images or an image they previously created/uploaded to the system). 
To guarantee a fair use of the testbed, Colosseum implements a token-based resource allocation system, where occupying resources costs a certain amount of tokens per hour.\footnote{Colosseum users are divided into teams, i.e., sets of users belonging to the same research group or organization. Each team is provided with a token budget that automatically resets on a weekly basis.}

After resources have been reserved, Colosseum automatically instantiates \gls{lxc} containers on the \glspl{srn}. Next, users log into Colosseum gateway through the Secure Shell (SSH) protocol and access the \glspl{srn} allocated to them. This is where the main part of Colosseum interactive experiments takes place. Users can start \gls{rf} and traffic scenarios by using \texttt{colosseumcli}---an \gls{api} to interface with some auxiliary services of Colosseum, e.g., \gls{mchem} and \gls{tgen}---and control the \glspl{usrp} X310 through softwarized protocol stacks instantiated on the \glspl{srn} (see Section~\ref{sec:usecases} for exemplary Colosseum use cases).
Through \texttt{colosseumcli}, users can also make a \textit{snapshot} of the container running on the \glspl{srn}. This creates an \gls{lxc} image from the container and saves it on Colosseum \gls{nas} for later use.
Finally, when the experiment ends, resources are deallocated and the content of the \texttt{/logs} directory of each container are saved on the \gls{nas}, from where they can be retrieved through the file proxy server.

\textbf{\textit{Batch Mode.}}
While the interactive mode is more suitable for designing, prototyping and troubleshooting solutions, the batch mode can be used to automatically run experiments in bulk, e.g., to benchmark solutions or perform data collection. 
Batch jobs are configured through \texttt{json}-formatted files, in which users can specify the experiment duration, which \gls{rf} and traffic scenarios to run, the number of \glspl{srn} to allocate and which \gls{lxc} images to instantiate on them. Additional parameters---handled by the user-defined programs---can also be passed.
Once scheduled, batch jobs are inserted in a queue and kicked off based on resource availability. Once a batch job starts, Colosseum takes care of instantiating the containers and to run the specified emulation scenario. Then, user-defined startup scripts are invoked on the \glspl{srn} and the actual experiment is carried out. Once the batch job ends, data and logs are copied from the \texttt{/logs} directory of the \gls{srn} containers to the \gls{nas}, analogously to what happens in interactive mode.

\section{Use Cases of Colosseum }
\label{sec:usecases}

This section illustrates a set of Colosseum use cases. Examples concerning cellular networking
are shown in Section~\ref{sec:cellular}, applications using a software-defined Wi-Fi protocol stack are showcased in Section~\ref{sec:wifi}, while spectrum sharing capabilities are discussed in Section~\ref{sec:spectrum-sharing}. Finally, the application of \glspl{uav} to wireless networking is demonstrated in Section~\ref{sec:uav}.

\subsection{Cellular Networking}
\label{sec:cellular}

\begin{figure}[ht]
    \centering
    \setlength\abovecaptionskip{-0pt}
    \setlength\fwidth{0.9\columnwidth}
    \setlength\fheight{0.25\columnwidth}
    \input{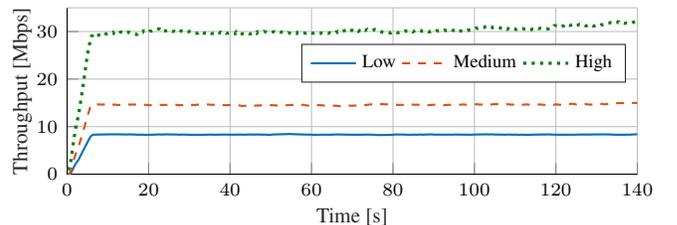}
    \caption{Downlink throughput with different traffic conditions.}
    \label{fig:cellular-thr}
\end{figure}

Softwarization and virtualization will play a fundamental role in 5G and beyond cellular networks. Telecom operators will deploy open source custom solutions on a ``white-box'' agnostic \gls{ran} where services, e.g., the base stations, will be instantiated on-demand~\cite{bonati2020open,doro2021coordinated}. This not only endows the network with flexibility by design, but it also allows real-time optimization of the users' service based on the current network conditions and traffic demand~\cite{bonati2020cellos, bonati2021intelligence}. Although open-source approaches bring unprecedented performance improvements, they require a reliable development environment and at scale evaluation
before they are deployed on the commercial infrastructure. More importantly, this testing phase needs to account for different environments, wireless channel and traffic conditions. With its unique emulation capabilities, Colosseum allows the research community to reliably prototype and benchmark solutions for future cellular networks.
To showcase these capabilities, we instantiate a cellular network with 4~base stations and 24~users through srsRAN~\cite{gomez2016srslte}. Figure~\ref{fig:cellular-thr} shows the network downlink throughput when the users request traffic through \texttt{iperf3} at different rates: low traffic ($1\:\mathrm{Mbps}$ per user), medium traffic ($2\:\mathrm{Mbps}$), and high traffic ($4\:\mathrm{Mbps}$). In this case, the base stations use a $10\:\mathrm{MHz}$ channel bandwidth.

\ifexttikz
    \tikzsetnextfilename{cellular-throughput-traffic}
\fi

\begin{figure}[ht]
    \centering
    \setlength\abovecaptionskip{-0pt}
    \ifoverleaf
    \else
        \ifexttikz
            \tikzsetnextfilename{cellular-speff-static}
        \fi
    \fi
    \begin{subfigure}[b]{0.48\columnwidth}
        \setlength\abovecaptionskip{-0pt}
        \ifoverleaf
            \includegraphics[width=\columnwidth]{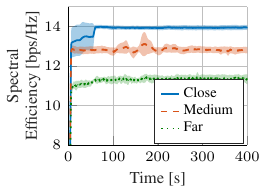}
        \else
            \setlength\fwidth{0.75\columnwidth}
            \setlength\fheight{0.55\columnwidth}
            \input{figures/cellular-speff-static.tex}
        \fi
        \caption{Varying distance.}
        \label{fig:cellular-speff-distance}
    \end{subfigure}
    \hfill
    \ifoverleaf
    \else
        \ifexttikz
            \tikzsetnextfilename{cellular-speff-mobility}
        \fi
    \fi
    \begin{subfigure}[b]{0.48\columnwidth}
        \setlength\abovecaptionskip{-0pt}
        \ifoverleaf
            \includegraphics[width=\columnwidth]{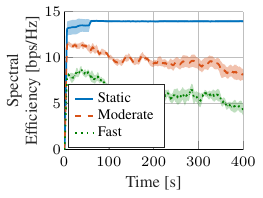}
        \else
            \setlength\fwidth{0.75\columnwidth}
            \setlength\fheight{0.55\columnwidth}
            \input{figures/cellular-speff-mobility.tex}
        \fi
        \caption{Varying speed.}
        \label{fig:cellular-speff-speed}
    \end{subfigure}
    \caption{Downlink spectral efficiency of the network varying the distance between users and base stations, and speed of the users.}
    \label{fig:cellular-speff}
    \vspace{-10pt}
\end{figure}

Figure~\ref{fig:cellular-speff} shows how Colosseum scenarios can be used to study the impact of deployment distance and user mobility in cellular networks~\cite{bonati2021scope}. We evaluate a network with 10~base stations and 40~users in close proximity to the base stations. Users request downlink video traffic at a rate of $1\:\mathrm{Mbps}$ through one of Colosseum traffic scenarios.
Specifically, Fig.~\ref{fig:cellular-speff-distance} shows the downlink spectral efficiency of the network for different deployment distances between users and base stations: close (users are deployed within $20\:\mathrm{m}$ from the base stations), medium ($50\:\mathrm{m}$), and far ($100\:\mathrm{m}$). Figure~\ref{fig:cellular-speff-speed}, instead, depicts the same metric for different configurations of user mobility: static (no mobility), moderate (users move at a speed of $3\:\mathrm{m/s}$), and fast ($5\:\mathrm{m/s}$). In this case, the users are deployed in close proximity to the base stations.
%
We notice that although the spectral efficiency decreases with the increased mobility/distance between users and base stations, the network performance shows tight $95$\% confidence intervals (shaded areas). This confirms the reliability and repeatability of Colosseum emulation.

\begin{figure}[ht]
    \centering
    \setlength\belowcaptionskip{-5pt}
    \includegraphics[width=\columnwidth]{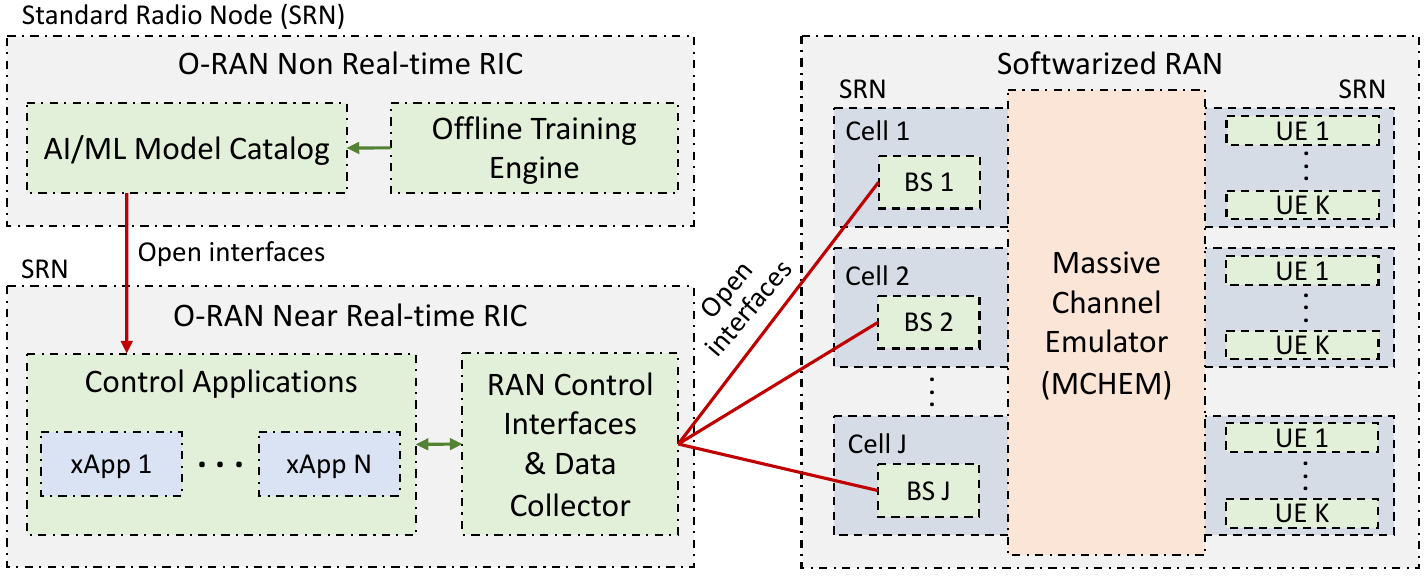}
    \caption{Example of O-RAN integration in Colosseum.}
    \label{fig:o-ran}
\end{figure}

\textbf{\textit{O-RAN.}}
Another use case of interest to cellular networks is that of the Open \gls{ran}, which envisions the deployment of open source and softwarized stacks and components on a ``white-box'' hardware infrastructure.
In this regard, the standardization efforts of the O-RAN Alliance are worth mentioning. This consortium is working at the standardization of open interfaces among the white-box network components---which would enable interoperability of equipment from multiple vendors---and at the creation of open controllers for the cellular infrastructure, namely \glspl{ric}.
These controllers operate at different timescales to govern the components of the Open \gls{ran} and enable the execution of third-party applications, e.g., xApps, that control the \gls{ran} elements (see Fig.~\ref{fig:o-ran})~\cite{bonati2020open, bonati2021intelligence}.


Based on the developments of~\cite{bonati2021intelligence}, we now showcase how Colosseum can be used to instantiate an O-RAN-managed cellular infrastructure. The performance of the network is optimized through data-driven closed-loop control enabled by xApps deployed at the near-real time \gls{ric}.
We deploy 4~softwarized base stations and 40~users
in a dense urban scenario.
The base stations divide the available spectrum in three slices---\gls{embb}, \gls{mtc}, and \gls{urllc}---and they run our SCOPE framework~\cite{bonati2021scope}, which enables dynamic control of slicing allocation and scheduling policies.
We control the configuration of each slice in real time through \gls{drl} agents
deployed as xApps on the near-real time \gls{ric}. The agents receive periodic data reports from the base stations and send them control directives through the O-RAN E2 termination. An example of these control directives is the selection of which scheduling policy among Round-robin (RR), Waterfilling (WF), and Proportionally Fair (PF) to implement for each network slice of each base stations.

\ifexttikz
    \tikzsetnextfilename{prb-allocation-ratio-slice-2}
\fi
\begin{figure}[ht]
    \centering
    \setlength\belowcaptionskip{-5pt}
    \setlength\fwidth{0.9\columnwidth}
    \setlength\fheight{0.3\columnwidth}
    \input{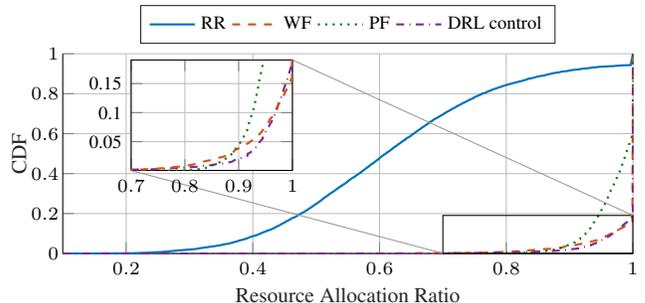}
    \caption{Resource allocation ratio of the URLLC slice for different scheduling policies and with DRL control.}
    \label{fig:oran-allocation-ratio}
\end{figure}

Figure~\ref{fig:oran-allocation-ratio} shows the \gls{cdf} of the resources allocated to the \gls{urllc} slice when the RR, WF, and PF scheduling policies are implemented statically, and with \gls{drl} control. For this slice, the agents aim at maximizing the ratio of resources granted to the users and those they request.
Since the \gls{drl} agents dynamically choose the optimal scheduling policy based on the current channel conditions and traffic demand, they achieve superior performance compared to the static scheduling policies.

\subsection{Wi-Fi}
\label{sec:wifi}
 
Wi-Fi has become an indispensable and ubiquitous wireless technology to provide home networking and public/hotspot Internet connectivity, as well as supporting a wide range of \gls{iot} applications.
%
To support the ever-increasing number of devices, the \gls{fcc} has recently opened an additional $1.2\:\mathrm{GHz}$ of spectrum in the $6\:\mathrm{GHz}$ band~\cite{fcc-6ghz}. This portion of the spectrum, intended for unlicensed use, has been embraced in the IEEE 802.11ax amendment, namely Wi-Fi~6~\cite{ieee-802_11ax-standard}.
Large-scale experimentation is a core enabler for research and development in these emerging spectrum bandwidths.

To showcase how Colosseum can facilitate these operations, we instantiate a Wi-Fi network through an open-source GNU Radio-based implementation of the IEEE~802.11a/g/p standard~\cite{bloessl2018performance}.
\ifexttikz
    \tikzsetnextfilename{wifi-snr-rome}
\fi
\begin{figure}[ht]
    \centering
    \setlength\abovecaptionskip{-0pt}
    \setlength\belowcaptionskip{-10pt}
    \setlength\fwidth{0.85\columnwidth}
    \setlength\fheight{0.2\columnwidth}
    \input{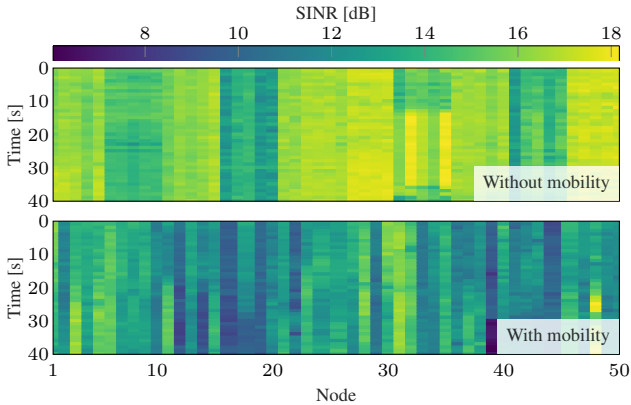}
    \caption{SINR measured by Wi-Fi nodes on Colosseum in a scenario without node mobility (top) and with node mobility (bottom).}
    \label{fig:wifi-sinr}
\end{figure}
We deploy 50~Wi-Fi transceivers in a Colosseum urban scenario and measure the \gls{sinr} in the case nodes do not move (Fig.~\ref{fig:wifi-sinr}, top), and in the case they move with an average speed of $3\:\mathrm{m/s}$ (Fig.~\ref{fig:wifi-sinr}, bottom).
We notice that when node mobility is implemented, the \gls{sinr} degrades by $3.18\:\mathrm{dB}$ on average because of the rapidly-changing channel conditions.

\subsection{Spectrum Sharing}
\label{sec:spectrum-sharing}

In recent years, the widespread use of wireless communications has fostered technological advancement and motivated the development and deployment of innovative services for mobile users. These phenomena have resulted in a sharp increase in the number of wireless devices and traffic demand. This has exacerbated the issue of the lack of spectrum to support them, also known as \textit{spectrum crunch}~\cite{yang2021spectrum,naik2021coexistance}.

In most cases, spectrum is either allocated to government agencies for research and military uses (e.g., radars, law enforcement), to the scientific community (e.g., for radio astronomy, atmospheric sensing), or licensed to operators, which gain exclusive access to spectrum frequencies in exchange for very high licensing fees.
Only in a handful of cases, such as the Industrial, Scientific and Medical (ISM) band, spectrum is allocated to non-commercial and experimental applications. One of the main drawbacks of such a static allocation is that unlicensed portions of the spectrum are extremely congested, while licensed ones might experience lower load conditions due to the lack of activity.

In this context, one research topic that has gained momentum
is spectrum sharing, whose ultimate goal is to devise solutions that allow multiple technologies to share portions of the spectrum.
%
To this end, \glspl{sdr} have gained increasing attention in the community as a promising solution to allow unlicensed users
to fill the so-called \textit{spectrum holes} left by licensed users.
%
Although the literature is characterized by several works that leverage optimization and data-driven solutions in \glspl{sdr}, one of the main limitations of such works is the lack of large-scale experimentation that demonstrates their effectiveness in heterogeneous network topologies, traffic and channel conditions~\cite{doostmohammady2021good,bonati2021scope}.

\begin{figure}[ht]
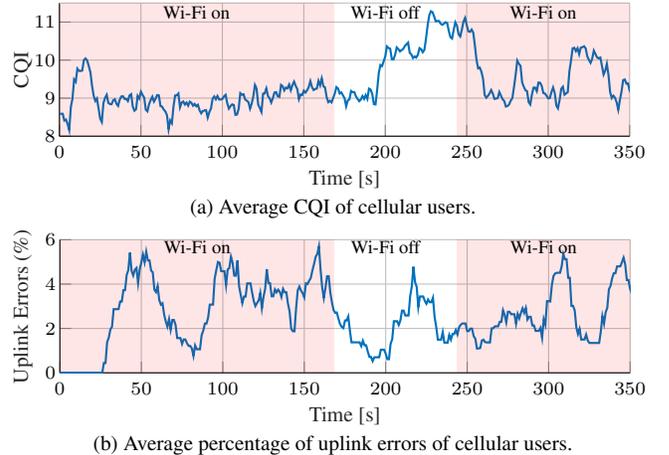

    \centering
    \setlength\abovecaptionskip{-0pt}
    \ifexttikz
        \tikzsetnextfilename{cell-cqi}
    \fi
    \begin{subfigure}[b]{\columnwidth}
        \centering
        \setlength\abovecaptionskip{-0pt}
        \setlength\fwidth{0.9\columnwidth}
        \setlength\fheight{0.2\columnwidth}
        \input{figures/cell-cqi.tex}
        \caption{Average CQI of cellular users.}
        \label{fig:spectrum-sharing-cqi}
    \end{subfigure}\\
    \ifexttikz
        \tikzsetnextfilename{cell-errors}
    \fi
    \begin{subfigure}[b]{\columnwidth}
        \centering
        \setlength\abovecaptionskip{-0pt}
        \setlength\fwidth{0.9\columnwidth}
        \setlength\fheight{0.2\columnwidth}
        \input{figures/cell-errors.tex}
        \caption{Average percentage of uplink errors of cellular users.}
        \label{fig:spectrum-sharing-errors}
    \end{subfigure}
    \caption{Average cell CQI and percentage of uplink errors with and without Wi-Fi traffic. Time periods without Wi-Fi traffic are marked with a red shaded area.}
    \label{fig:spectrum-sharing}
\end{figure}

Colosseum fills this gap and allows users to instantiate large-scale networks with nodes running heterogeneous wireless protocol stacks (e.g., cellular, Wi-Fi) on the same spectrum bandwidth. This enables researchers to design, implement and evaluate novel spectrum sharing solutions on heterogeneous and diverse network deployments and conditions without causing harmful interference to licensed users.
%
%
To provide an example of this application, we deploy on the same portion of the spectrum a cellular base station serving 5~cellular users, and two Wi-Fi nodes.
%
Figure~\ref{fig:spectrum-sharing} shows the average \gls{cqi} and percentage of uplink errors experienced by the cellular users with and without Wi-Fi transmissions active.
We notice that the interference generated by Wi-Fi communications causes the \gls{cqi} of the cellular users to rapidly drop (Fig.~\ref{fig:spectrum-sharing-cqi}), and the percentage of uplink errors to increase (Fig.~\ref{fig:spectrum-sharing-errors}) when Wi-Fi transmissions are ongoing.

\subsection{Unmanned Aerial Vehicles}
\label{sec:uav}

The growing commercialization of \glspl{uav}, along with their greater affordability and increasing popularity, has recently spawned the interest of telecom operators and equipment providers toward the use of drones for networking applications.
%
Examples of such applications include the use of \glspl{uav} to create swarm of flying networks~\cite{mohanti2019airbeam,bertizzolo2020swarmcontrol}, to promptly aid in the aftermath of disaster scenarios~\cite{coletta2020danger,ferranti2021hironet,rottondi2021scheduling}, and to bring connectivity to remote areas~\cite{bertizzolo2019mmbac,ferranti2020skycell}.
%
However, security concerns, strict flying regulations, and the cost of the equipment itself impose additional challenges to the development and testing of solutions at scale.

\begin{figure}[htb]
    \centering
    \setlength\belowcaptionskip{-5pt}
    \begin{subfigure}[b]{0.45\columnwidth}
        \setlength\abovecaptionskip{20pt}
        \includegraphics[trim={1.7cm 1.7cm 2.5cm 1.7cm},clip,width=\columnwidth,height=2.5cm]{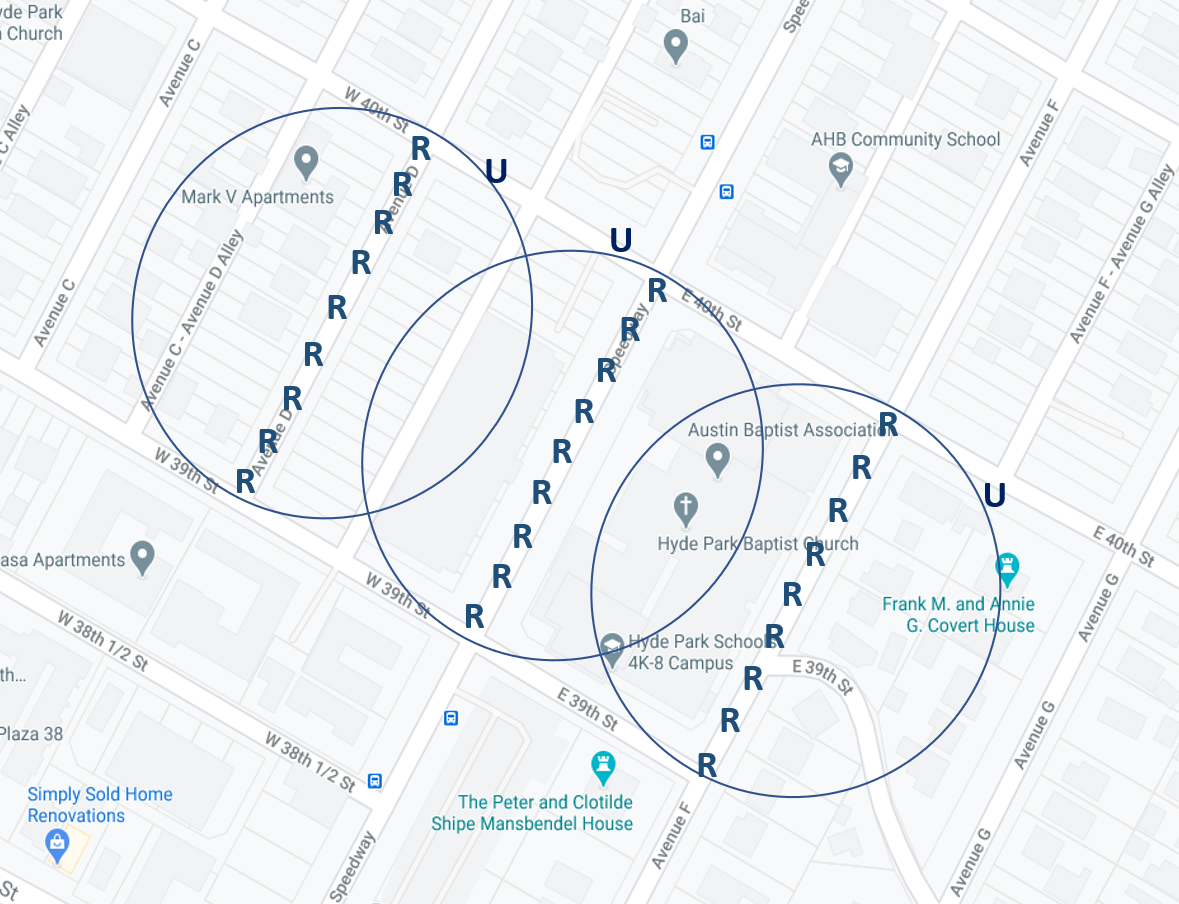}
        \caption{Network topology.}
        \label{fig:aoatopo}
    \end{subfigure}
    \ifexttikz
        \tikzsetnextfilename{uav-austin-squad-sinr}
    \fi
    \begin{subfigure}[b]{0.51\columnwidth}
        \setlength\abovecaptionskip{20pt}
        \setlength\fwidth{3.8cm}
        \setlength\fheight{2.5cm}
        \hspace{-2.7cm}\vspace{-0.7cm}
%
%
\definecolor{mycolor1}{rgb}{0.00000,0.44700,0.74100}%
\definecolor{mycolor2}{rgb}{0.85000,0.32500,0.09800}%
\definecolor{mycolor3}{rgb}{0.92900,0.69400,0.12500}%
\definecolor{mycolor4}{rgb}{0.49400,0.18400,0.55600}%
\definecolor{mycolor5}{rgb}{0.46600,0.67400,0.18800}%
\definecolor{mycolor6}{rgb}{0.30100,0.74500,0.93300}%
\definecolor{mycolor7}{rgb}{0.302, 0.686, 0.29}%
\begin{tikzpicture}
\pgfplotsset{every tick label/.append style={font=\scriptsize}}

\tikzset{
    hatch distance/.store in=\hatchdistance,
    hatch distance=2.5pt,
    hatch thickness/.store in=\hatchthickness,
    hatch thickness=0.5pt
}
    
 \makeatletter
\pgfdeclarepatternformonly[\hatchdistance,\hatchthickness]{flexible hatch}
{\pgfqpoint{0pt}{0pt}}
{\pgfqpoint{\hatchdistance}{\hatchdistance}}
{\pgfpoint{\hatchdistance}{\hatchdistance}}%
{
    \pgfsetcolor{\tikz@pattern@color}
    \pgfsetlinewidth{\hatchthickness}
    \pgfpathmoveto{\pgfqpoint{0pt}{0pt}}
    \pgfpathlineto{\pgfqpoint{\hatchdistance}{\hatchdistance}}
    \pgfusepath{stroke}
}
\makeatother

\pgfmathsetlengthmacro\MajorTickLength{
    \pgfkeysvalueof{/pgfplots/major tick length} * 0.75
}

\begin{axis}[%
width=0.951\fwidth,
height=\fheight,
at={(0\fwidth,0\fheight)},
scale only axis,
xtick={1,2,3},
xlabel style={font=\footnotesize\color{white!15!black}},
xlabel={Squad},
ymin=-2,
ymax=14,
ytick={-2,0,5,10},
extra x ticks={3.5},
extra x tick labels={},
extra x tick style={tick style={draw=none}},
extra y ticks={14},
extra y tick labels={},
extra y tick style={tick style={draw=none}},
ylabel style={font=\footnotesize\color{white!15!black}},
ylabel={SINR [dB]},
axis background/.style={fill=white},
axis x line*=bottom,
axis y line*=left,
xmajorgrids,
ymajorgrids,
legend style={legend cell align=left, align=left, draw=white!15!black, anchor=south, font=\scriptsize, at={(-0.1,1.05)}, legend style={row sep=-2pt}},
legend columns=2,
ylabel shift=-9pt,
xlabel shift=-5pt,
enlarge x limits={abs=0.5},
ybar,
bar width=10pt,
xtick align=inside,
major tick length=\MajorTickLength,
]

\addplot[fill=mycolor1!50, draw=mycolor1, area legend]
  table[row sep=crcr] {%
1  12.7064\\
2  9.0143\\
3  12.1342\\
};
\addlegendentry{Max weighted matching}

\addplot[fill=mycolor2!50, draw=mycolor2, area legend, pattern=flexible hatch, pattern color=mycolor2]
  table[row sep=crcr] {%
1  0.491\\
2  -1.6651\\
3  1.046\\
};
\addlegendentry{Random allocation}


\end{axis}
\end{tikzpicture}%
        \caption{Average squad SINR.}
        \label{fig:aoasinr}
    \end{subfigure}
    \caption{Left: network topology of an experiment with 30~nodes (ground relays, ``R'', and UAVs, ``U'') in the Alleys of Austin scenario. The circles mark the trajectory of the UAVs. Right: average SINR of three squads in Alleys of Austin scenario when different algorithms are used for the selection of relay nodes.}
    \label{fig:uav-alleys}
    \vspace{-5pt}
\end{figure}
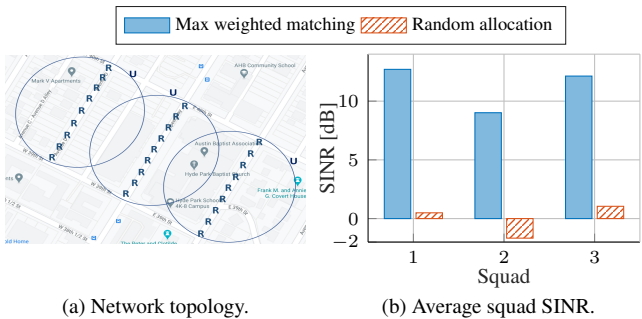

Colosseum has the potential of facilitating the evaluation of such solutions by providing the means to emulate large-scale \gls{sdr}-enabled \gls{uav} networks.
To showcase these capabilities, we instantiate a network with 30~nodes in the Alleys of Austin scenario (see Section~\ref{sec:scenario-samples}).
Nodes are divided in three groups, or \textit{squads} (see Fig.~\ref{fig:aoatopo}). Each squad is composed of a \gls{uav} (marked with ``U'' in the figure) and 9~ground relay nodes (``R''), which need to deliver data (e.g., video and voice traffic) to specific members of the squad.

At each instant of time, specific nodes act as relays (aerial or ground) for data from a squad to the intended destination (e.g., nodes in other squads), while coping with inter-user and inter-squad interference.
Figure~\ref{fig:aoasinr} shows the average \gls{sinr} when two different algorithms are used to select the relay nodes: \gls{sinr}-based maximum weighted matching,
and random allocation.
%
We notice that when the maximum weighted matching algorithm is used, the squads experience a higher \gls{sinr} than when the random allocation strategy is enforced.
This is due to the fact that the former algorithm elects as relays those nodes that have better channel conditions (and thus a higher \gls{sinr}) toward the remaining squad members. This algorithm also takes advantage of the \glspl{uav} as aerial relays, which are in line-of-sight condition with the ground nodes.

\section{Evolution of Colosseum}
\label{sec:evolution}

While Colosseum represents an invaluable tool for the evaluation of large-scale wireless networking scenarios, it also needs to evolve to match the needs of communication technologies in the years to come. The following sections describe planned extensions of Colosseum that will add new emulation capabilities (Section~\ref{sec:nrdz}) and \gls{ai} components (Section~\ref{sec:aijump}).

\subsection{Expansions to Support NRDZ and mmWave Research}
\label{sec:nrdz}

A steady trend in the wireless industry has been the push for the allocation of additional spectrum for communications, with spectrum sharing~\cite{zhang2017survey} and the adoption of new frequency bands, e.g., \acrlongpl{mmwave}~\cite{rangan2014millimeter,giordani2018tutorial}. As the usage of the spectrum expands into new frontiers, it becomes important to study and understand how different spectrum users can safely coexist. Along this line, the \gls{nsf} has recently published a ``Dear Colleague Letter'' to explore the feasibility of establishing \glspl{nrdz} in the United States. \glspl{nrdz} are meant as safe platforms for experiments that would not be allowed under \gls{fcc} spectrum regulations because of chances to generate harmful interference toward incumbent users. Colosseum is naturally positioned to implement an \gls{nrdz}, as---by being an emulator---it avoids any interference to external devices. Additionally, its programmability allows experimenters to test multiple scenarios, different protocol stacks, and different traffic patterns (see Section~\ref{sec:scenarios}).

\begin{figure}
    \centering
    \setlength\belowcaptionskip{-15pt}
    \includegraphics[width=\columnwidth]{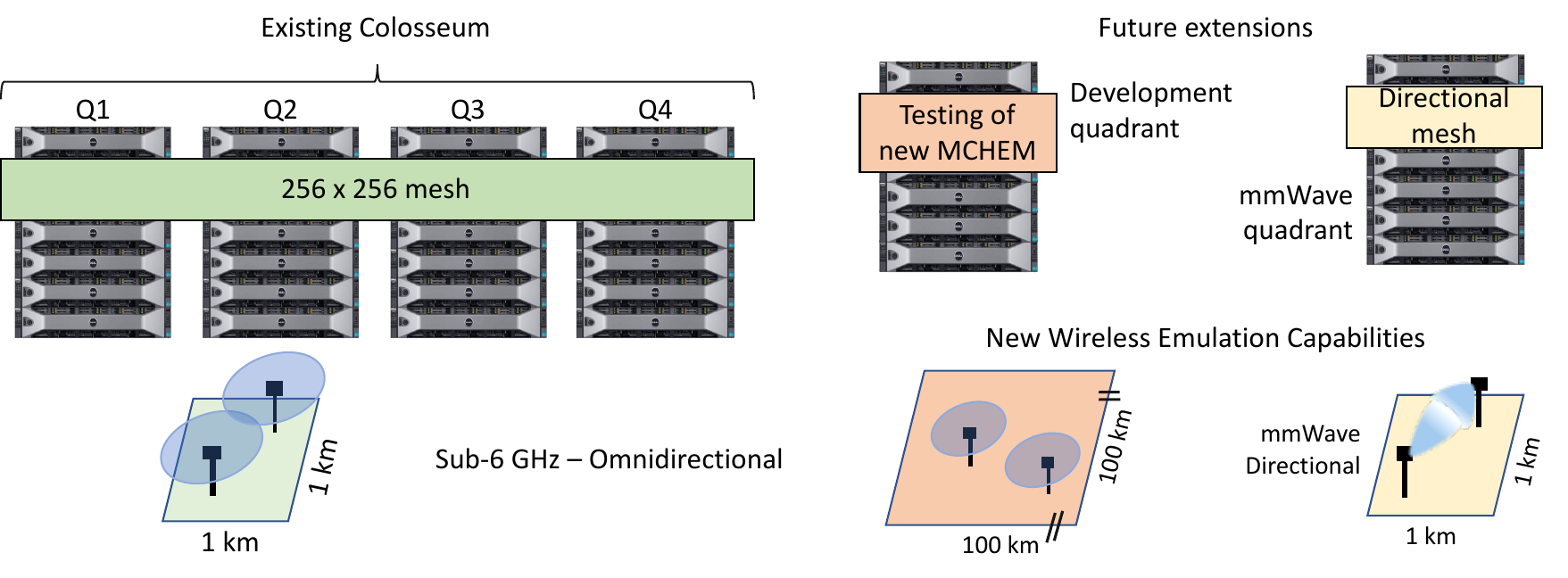}
    \caption{Current components of Colosseum and proposed extensions for the NRDZ program.}
    \label{fig:colosseum-new}
\end{figure}

As part of the \gls{nrdz} activities, Colosseum will be expanded to support new spectrum bands, scenarios, and power levels. Colosseum already supports very granular coordinated spectrum sharing across heterogeneous networks,  and will act as a ``management and optimizer'' system for protection, observation, validation and automation of spectrum sharing. These capabilities will be extended by adding two new quadrants (see Fig.~\ref{fig:colosseum-new}) and by improving the current \gls{mchem} emulation. 

As of today, there are two limitations in Colosseum that prevent it to act as a tool to design and study NRDZs as defined above. First, Colosseum was designed to emulate terrains up to $1\!\times\!1\:\mathrm{km}$. This limitation is a consequence of the maximum propagation delay that can be emulated by \gls{mchem}, which is embedded in the design of its \glspl{fpga}. Second, Colosseum was designed to emulate sub-$6\:\mathrm{GHz}$ omnidirectional transmissions. However, it cannot emulate directional transmissions and higher frequency bands.

To go beyond today's limitations, we are deploying an additional, reduced version of a Colosseum quadrant as a development environment. This environment includes a dedicated \gls{mchem} quadrant that will be the basis for the development of new channel emulation code, as well as the testing of new hardware for the user radios and \glspl{srn}.
Notably, two extensions for \gls{mchem} will be introduced first in this environment, and then scaled to the full Colosseum (Fig.~\ref{fig:colosseum-new}). The first will allow longer delays for high-fidelity emulation of a $100\times100\:\mathrm{km}$ grid and terrains representative of \glspl{nrdz}. The second will introduce the modeling of beamforming-based, directional communications to emulate mmWave networks.
Finally, to support the bandwidth requirements of 5G and beyond cellular networks, we will create a new Colosseum quadrant able to emulate very large baseband bandwidth (up to $2\:\mathrm{GHz}$ of bandwidth).


\subsection{AI Jumpstart Integration}
\label{sec:aijump}

Colosseum is currently being expanded through the MassTech ``\gls{ai} Jumpstart'' program~\cite{ai-jumpstart}.
This program aims at jump-starting Massachusetts firms interested in deploying \gls{ai} to enhance their businesses, connecting industry practitioners with world-class facilities and researchers.
Part of the \gls{ai} Jumpstart equipment will be integrated in the Colosseum environment to enable research and development of \gls{ai}-based wireless networking solutions. Examples of use cases are the fast and efficient training of large-scale wireless datasets collected on Colosseum~\cite{bonati2021intelligence}; real-time, AI-driven 5G signal processing for the full 5G protocol stack~\cite{polese2021deepbeam}; and model-free adaptation and control of large-scale wireless networks.

The new equipment includes two NVIDIA DGX A100 nodes---among the most powerful \gls{ai} compute solutions on the market today---capable of delivering up to $10\:\mathrm{petaFLOPS}$~\cite{a100}. These nodes include 8~GPUs, state-of-the-art AMD CPUs, $1\:\mathrm{TB}$ of RAM, and 10~Mellanox ConnectX-6 network cards, each capable of sustaining a $200\:\mathrm{Gbps}$ link. An additional large memory node enables memory-intensive workloads, with $3\:\mathrm{TB}$ of RAM. These machines are connected through a dedicated Mellanox Infiniband switch, which can sustain an aggregated traffic of up to $16\:\mathrm{Tbps}$. This system will be fully meshed with the compute and wireless resources of Colosseum, with dedicated Nomad-based orchestration and load balancing capabilities~\cite{nomad}.

\section{Conclusions}
\label{sec:conclusions}

This paper presents Colosseum---the world's largest wireless network emulator with hardware-in-the-loop---as a testbed that is \textit{for the first time} publicly available to the community.
Colosseum
enables experimental research at scale through virtualized and softwarized waveforms and protocol stacks.
The unique emulation capabilities of Colosseum enable the design, prototyping, and testing of wireless solutions in a host of scenarios and channel conditions.
We illustrate the architecture, emulation capabilities and operational modes of Colosseum. 
We then provide examples of large-scale experimentation in Colosseum, considering cellular, Wi-Fi, spectrum sharing and \gls{uav} scenarios.
We finally lay out a plan for extending Colosseum to enable experiments over larger deployments and higher frequencies.

\balance
\footnotesize  
\bibliographystyle{IEEEtran}
\bibliography{biblio}

\begin{thebibliography}{10}
\providecommand{\url}[1]{#1}
\csname url@samestyle\endcsname
\providecommand{\newblock}{\relax}
\providecommand{\bibinfo}[2]{#2}
\providecommand{\BIBentrySTDinterwordspacing}{\spaceskip=0pt\relax}
\providecommand{\BIBentryALTinterwordstretchfactor}{4}
\providecommand{\BIBentryALTinterwordspacing}{\spaceskip=\fontdimen2\font plus
\BIBentryALTinterwordstretchfactor\fontdimen3\font minus
  \fontdimen4\font\relax}
\providecommand{\BIBforeignlanguage}[2]{{%
\expandafter\ifx\csname l@#1\endcsname\relax
\typeout{** WARNING: IEEEtran.bst: No hyphenation pattern has been}%
\typeout{** loaded for the language `#1'. Using the pattern for}%
\typeout{** the default language instead.}%
\else
\language=\csname l@#1\endcsname
\fi
#2}}
\providecommand{\BIBdecl}{\relax}
\BIBdecl

\bibitem{bonati2020open}
L.~Bonati, M.~Polese, S.~D’Oro, S.~Basagni, and T.~Melodia, ``{Open,
  Programmable, and Virtualized 5G Networks: State-of-the-Art and the Road
  Ahead},'' \emph{Computer Networks}, vol. 182, pp. 1--18, December 2020.

\bibitem{naik2021coexistance}
G.~Naik and J.-M.~J. Park, ``{Coexistence of Wi-Fi 6E and 5G NR-U: Can We Do
  Better in the 6 GHz Bands?}'' in \emph{Proceedings of IEEE INFOCOM},
  Vancouver, BC, Canada, May 2021.

\bibitem{bertizzolo2020swarmcontrol}
L.~Bertizzolo \emph{et~al.}, ``{SwarmControl: An Automated Distributed Control
  Framework for Self-Optimizing Drone Networks},'' in \emph{Proceedings of IEEE
  INFOCOM}, Toronto, ON, Canada, July 2020.

\bibitem{BuczekBBM21}
J.~Buczek, L.~Bertizzolo, S.~Basagni, and T.~Melodia, ``\emph{What is A
  Wireless {UAV}?} a design blueprint for {6G} flying wireless nodes,'' in
  \emph{Proceedings of ACM WiNTECH 2021}, New Orleans, LA, January 2022.

\bibitem{santagati2017softwaredefined}
G.~E. Santagati and T.~Melodia, ``{A Software-Defined Ultrasonic Networking
  Framework for Wearable Devices},'' \emph{IEEE/ACM Transactions on
  Networking}, vol.~25, no.~2, pp. 960--973, April 2017.

\bibitem{breen2021powder}
J.~Breen \emph{et~al.}, ``{POWDER: Platform for Open Wireless Data-driven
  Experimental Research},'' \emph{Computer Networks}, vol. 197, pp. 1--18,
  October 2021.

\bibitem{doostmohammady2021good}
R.~Doost-Mohammady, O.~Bejarano, and A.~Sabharwal, ``{Good Times for Wireless
  Research},'' \emph{Computer Networks}, vol. 188, pp. 1--9, April 2021.

\bibitem{kohli2021openaccess}
M.~Kohli \emph{et~al.}, ``{Open-Access Full-Duplex Wireless in the ORBIT and
  COSMOS Testbeds},'' \emph{Computer Networks}, 2021.

\bibitem{panicker2021aerpaw}
A.~Panicker \emph{et~al.}, ``{AERPAW Emulation Overview and Preliminary
  Performance Evaluation},'' \emph{Computer Networks}, vol. 194, pp. 1--11,
  July 2021.

\bibitem{zhang2021ara}
H.~Zhang \emph{et~al.}, ``{ARA: A Wireless Living Lab Vision for Smart and
  Connected Rural Communities},'' in \emph{Proceedings of ACM WiNTECH}, New
  Orleans, LA, USA, October 2021.

\bibitem{koumaras20185genesis}
H.~Koumaras \emph{et~al.}, ``{5GENESIS: The Genesis of a Flexible 5G
  Facility},'' in \emph{Proceedings of IEEE CAMAD}, Barcelona, Spain, September
  2018.

\bibitem{dandekar2019grid}
K.~R. Dandekar \emph{et~al.}, ``{Grid Software Defined Radio Network Testbed
  for Hybrid Measurement and Emulation},'' in \emph{Proceedings of IEEE SECON},
  Boston, MA, USA, June 2019.

\bibitem{bertizzolo2020arena}
L.~Bertizzolo \emph{et~al.}, ``{{Arena}: A 64-antenna {SDR}-based Ceiling Grid
  Testing Platform for Sub-6 {GHz} {5G}-and-Beyond Radio Spectrum Research},''
  \emph{Computer Networks}, pp. 1--17, November 2020.

\bibitem{darpa_sc2}
{DARPA}. {Spectrum Collaboration Challenge}.
  \url{https://www.darpa.mil/program/spectrum-collaboration-challenge}.
  Accessed September 2021.

\bibitem{tilghman2019will}
P.~Tilghman, ``{AI Will Rule the Airwaves: A DARPA Grand Challenge Seeks
  Autonomous Radios to Manage the Wireless Spectrum},'' \emph{IEEE Spectrum},
  vol.~56, no.~6, pp. 28--33, May 2019.

\bibitem{yuan2019defense}
R.~L. Yuan and K.~M. Schmidt, ``{Defense Advanced Research Projects Agency
  Spectrum Collaboration Challenge at APL: Introduction},'' vol.~35, no.~1, pp.
  2--3, 2019.

\bibitem{coleman2019overview}
D.~Coleman \emph{et~al.}, ``{Overview of the Colosseum: The World’s Largest
  Test Bed for Radio Experiments},'' \emph{Johns Hopkins APL Technical Digest},
  vol.~35, no.~1, pp. 4--11, 2019.

\bibitem{freeman2019software}
A.~S. Freeman \emph{et~al.}, ``{Software Project Management for the Defense
  Advanced Research Projects Agency Spectrum Collaboration Challenge},''
  \emph{Johns Hopkins APL Technical Digest}, vol.~35, no.~1, pp. 12--21, 2019.

\bibitem{plummer2019development}
A.~T. Plummer, Jr. and K.~P. Taylor, ``{Development and Operations on the
  Defense Advanced Research Project Agency’s Spectrum Collaboration
  Challenge},'' \emph{Johns Hopkins APL Technical Digest}, vol.~35, no.~1, pp.
  22--33, 2019.

\bibitem{mok2019resource}
J.~W. Mok, A.~L. Hom, J.~J. Uher, and D.~M. Coleman, ``{The Resource Manager
  for the Defense Advanced Research Projects Agency Spectrum Collaboration
  Challenge Test Bed},'' \emph{Johns Hopkins APL Technical Digest}, vol.~35,
  no.~1, pp. 34--41, 2019.

\bibitem{white2019standard}
D.~A. White, Jr., J.~E. Annis, and F.~F. Johnson, ``{Standard Radio Nodes in
  the Defense Advanced Research Projects Agency Spectrum Collaboration
  Challenge},'' vol.~35, no.~1, pp. 42--48, 2019.

\bibitem{yim2019incumbent}
K.~J. Yim, K.~R. McKeever, and D.~R. Barcklow, ``{Incumbent Radio Systems in
  the Defense Advanced Research Projects Agency Spectrum Collaboration
  Challenge Test Bed},'' vol.~35, no.~1, pp. 49--57, 2019.

\bibitem{curtis2019traffic}
P.~D. Curtis, A.~T. Plummer, Jr., J.~E. Annis, and W.~J. La~Cholter, ``{Traffic
  Generation System for the Defense Advanced Research Projects Agency Spectrum
  Collaboration Challenge},'' \emph{Johns Hopkins APL Technical Digest},
  vol.~35, no.~1, pp. 58--68, 2019.

\bibitem{barcklow2019radio}
D.~Barcklow \emph{et~al.}, ``{Radio Frequency Emulation System for the Defense
  Advanced Research Projects Agency Spectrum Collaboration Challenge},''
  \emph{Johns Hopkins APL Technical Digest}, vol.~35, no.~1, pp. 69--78, 2019.

\bibitem{bonati2021scope}
L.~Bonati, S.~D'Oro, S.~Basagni, and T.~Melodia, ``{{SCOPE}: An Open and
  Softwarized Prototyping Platform for {NextG} Systems},'' in \emph{Proceedings
  of ACM MobiSys}, Virtual Conference, June 2021.

\bibitem{bonati2021intelligence}
L.~Bonati, S.~D'Oro, M.~Polese, S.~Basagni, and T.~Melodia, ``{Intelligence and
  Learning in {O-RAN} for Data-driven {NextG} Cellular Networks},'' \emph{IEEE
  Communications Magazine}, 2021.

\bibitem{ashish2018scalable}
A.~Chaudhari and M.~Braun, ``{A Scalable FPGA Architecture for Flexible,
  Large-Scale, Real-Time RF Channel Emulation},'' in \emph{Proceedings of IEEE
  ReCoSoC}, Lille, France, July 2018.

\bibitem{mgen}
{U.S. Naval Research Laboratory}. {MGEN Traffic Emulator}.
  \url{https://www.nrl.navy.mil/Our-Work/Areas-of-Research/Information-Technology/NCS/MGEN}.
  Accessed September 2021.

\bibitem{polese2020experimental}
M.~Polese, L.~Bertizzolo, L.~Bonati, A.~Gosain, and T.~Melodia, ``{An
  Experimental mmWave Channel Model for UAV-to-UAV Communications},'' in
  \emph{Proceedings of ACM mmNets}, London, United Kingdom, September 2020.

\bibitem{tehranimoayyed2021creating}
M.~Tehrani-Moayyed, L.~Bonati, P.~Johari, T.~Melodia, and S.~Basagni,
  ``{Creating RF Scenarios for Large-Scale, Real-Time Wireless Channel
  Emulators},'' in \emph{Proceedings of IEEE MedComNet}, Ibiza, Spain, June
  2021.

\bibitem{opencellid}
\BIBentryALTinterwordspacing
{Unwired Labs}, ``{OpenCelliD},'' accessed September 2021. [Online]. Available:
  \url{https://opencellid.org}
\BIBentrySTDinterwordspacing

\bibitem{doro2021coordinated}
S.~D’Oro, L.~Bonati, F.~Restuccia, and T.~Melodia, ``{Coordinated 5G Network
  Slicing: How Constructive Interference Can Boost Network Throughput},''
  \emph{IEEE/ACM Transactions on Networking}, vol.~29, no.~4, pp. 1881--1894,
  August 2021.

\bibitem{bonati2020cellos}
L.~Bonati \emph{et~al.}, ``{{CellOS}: Zero-touch Softwarized Open Cellular
  Networks},'' \emph{Computer Networks}, vol. 180, pp. 1--13, October 2020.

\bibitem{gomez2016srslte}
I.~Gomez-Miguelez \emph{et~al.}, ``{{srsLTE}: An Open-source Platform for {LTE}
  Evolution and Experimentation},'' in \emph{Proceedings of ACM WiNTECH}, New
  York City, NY, USA, October 2016.

\bibitem{fcc-6ghz}
{FCC}. (2020, April) {FCC Adopts New Rules for the 6 GHz Band, Unleashing 1,200
  MHz of Spectrum for Unlicensed Use}.
  \url{https://docs.fcc.gov/public/attachments/DOC-363945A1.pdf}. Accessed
  September 2021.

\bibitem{ieee-802_11ax-standard}
{IEEE 802.11 Wireless LAN Working Group}, ``{IEEE Standard for Information
  Technology--Telecommunications and Information Exchange between Systems Local
  and Metropolitan Area Networks--Specific Requirements Part 11: Wireless LAN
  Medium Access Control (MAC) and Physical Layer (PHY) Specifications Amendment
  1: Enhancements for High-Efficiency WLAN},'' \emph{IEEE Std 802.11ax-2021
  (Amendment to IEEE Std 802.11-2020)}, pp. 1--767, May 2021.

\bibitem{bloessl2018performance}
B.~Bloessl, M.~Segata, C.~Sommer, and F.~Dressler, ``{Performance Assessment of
  IEEE 802.11p with an Open Source SDR-based Prototype},'' \emph{IEEE
  Transactions on Mobile Computing}, vol.~17, no.~5, pp. 1162--1175, May 2018.

\bibitem{yang2021spectrum}
P.~Yang, L.~Kong, and G.~Chen, ``{Spectrum Sharing for 5G/6G URLLC: Research
  Frontiers and Standards},'' \emph{IEEE Communications Standards Magazine},
  vol.~5, no.~2, pp. 120--125, June 2021.

\bibitem{mohanti2019airbeam}
S.~Mohanti \emph{et~al.}, ``{AirBeam: Experimental Demonstration of Distributed
  Beamforming by a Swarm of UAVs},'' in \emph{Proceedings of IEEE MASS},
  Monterey, CA, USA, November 2019.

\bibitem{coletta2020danger}
A.~Coletta, G.~Maselli, M.~Piva, and D.~Silvestri, ``{DANGER: A Drones Aided
  Network for Guiding Emergency and Rescue Operations},'' in \emph{Proceedings
  of ACM Mobihoc}, Virtual Conference, October 2020.

\bibitem{ferranti2021hironet}
L.~Ferranti, S.~D'Oro, L.~Bonati, F.~Cuomo, and T.~Melodia, ``{HIRO-NET:
  Heterogeneous Intelligent RObotic Network for Internet sharing in Disaster
  Scenarios},'' \emph{IEEE Transactions on Mobile Computing}, 2021.

\bibitem{rottondi2021scheduling}
C.~Rottondi, F.~Malandrino, A.~Bianco, C.~F. Chiasserini, and I.~Stavrakakis,
  ``{Scheduling of Emergency Tasks for Multiservice UAVs in Post-disaster
  Scenarios},'' \emph{Computer Networks}, vol. 184, pp. 1--13, January 2021.

\bibitem{bertizzolo2019mmbac}
L.~Bertizzolo \emph{et~al.}, ``{mmBAC: Location- aided mmWave Backhaul
  Management for UAV-based Aerial Cells},'' in \emph{Proceedings of ACM
  mmNets}, Los Cabos, Mexico, October 2019.

\bibitem{ferranti2020skycell}
L.~Ferranti, L.~Bonati, S.~D'Oro, and T.~Melodia, ``{SkyCell: A Prototyping
  Platform for 5G Aerial Base Stations},'' in \emph{Proceedings of IEEE
  SwarmNet}, Cork, Ireland, August 2020.

\bibitem{zhang2017survey}
L.~Zhang \emph{et~al.}, ``{A Survey of Advanced Techniques for Spectrum Sharing
  in {5G} Networks},'' \emph{IEEE Wireless Communications}, vol.~24, no.~5, pp.
  44--51, October 2017.

\bibitem{rangan2014millimeter}
S.~Rangan, T.~S. Rappaport, and E.~Erkip, ``{Millimeter-Wave Cellular Wireless
  Networks: Potentials and Challenges},'' \emph{Proceedings of the IEEE}, vol.
  102, no.~3, pp. 366--385, March 2014.

\bibitem{giordani2018tutorial}
M.~{Giordani}, M.~{Polese}, A.~{Roy}, D.~{Castor}, and M.~{Zorzi}, ``{A
  Tutorial on Beam Management for {3GPP NR} at {mmWave} Frequencies},''
  \emph{IEEE Communications Surveys \& Tutorials}, vol.~21, no.~1, pp.
  173--196, February 2019.

\bibitem{ai-jumpstart}
{The Innovation Institute at the MassTech Collaborative}. (2020, February) {AI
  Jumpstart}.
  \url{https://innovation.masstech.org/projects-and-initiatives/collaborative-research-matching-grant-program/ai-jumpstart}.
  Accessed September 2021.

\bibitem{polese2021deepbeam}
M.~Polese, F.~Restuccia, and T.~Melodia, ``{DeepBeam: Deep Waveform Learning
  for Coordination-Free Beam Management in mmWave Networks},'' in
  \emph{Proceedings of ACM Mobihoc}, Shanghai, China, July 2021.

\bibitem{a100}
NVIDIA. {NVIDIA DGX A100---The Universal System for AI Infrastructure}.
  \url{https://www.nvidia.com/en-us/data-center/dgx-a100}. Accessed September
  2021.

\bibitem{nomad}
{HashiCorp}. {Nomad}. \url{https://www.nomadproject.io}. Accessed September
  2021.

\end{thebibliography}

\end{document}